\documentclass[10pt, conference, letterpaper]{IEEEtran}

\usepackage{amsmath,amssymb,amsthm,array,color,colortbl,graphicx,multirow,xcolor}
\usepackage{microtype}

\usepackage{tikz}
\usetikzlibrary{decorations.pathmorphing, positioning,arrows.meta}

\usepackage{varwidth}
\usepackage{balance}

\usepackage[noend]{algorithm}
\usepackage[noend]{algpseudocode}

\usepackage[sort]{cite}
\usepackage{multirow}
\usepackage{colortbl}
\usepackage{url}

\usepackage{hhline}
\usepackage[multiple,para]{footmisc}

\newtheorem{definition}{Definition}
\newtheorem{theorem}{Theorem}

\newtheorem{corollary}{Corollary}
\newtheorem{lemma}{Lemma}

\definecolor{lgray}{rgb}{0.95,0.95,0.95}

\usepackage{cleveref}

\title{On the Price of Locality in Static Fast Rerouting}

\author{Klaus-Tycho Foerster$^*$ \quad Juho Hirvonen$^\dagger$ \quad Yvonne-Anne Pignolet$^\ddagger$ \quad Stefan Schmid$^\spadesuit$ \quad Gilles Tredan$^\diamondsuit$\\ $^*$TU Dortmund, Germany \quad $^\dagger$Aalto University, Finland \quad $^\ddagger$DFINITY, Switzerland\\ $^\spadesuit$TU Berlin, Germany \& University of Vienna, Austria \quad $^\diamondsuit$LAAS-CNRS, France}

\begin{document}

\maketitle

\thispagestyle{plain}
\pagestyle{plain}

\begin{abstract}
Modern communication networks feature fully decentralized flow rerouting mechanisms which allow them to quickly react to link failures.
This paper revisits the fundamental algorithmic problem underlying such local fast rerouting mechanisms.
%given a network $G=(V,E)$, compute for each node $v\in V$ a static routing table which, given the header of an incoming packet, specifies to which incident port the packet should be forwarded.
%As the routing tables need to be precomputed without knowledge of the actual failure scenario, they can only depend on \emph{local} failure information.
%In particular, this model raises the question whether it is possible to achieve \emph{perfect resilience} under multiple link failures: 
Is it possible to achieve \emph{perfect resilience}, i.e., to define local routing tables which  preserve connectivity as long as the underlying network is still connected?
Feigenbaum et al.~\cite{podc-ba} and Foerster et al.~\cite{apocs21resilience} showed that, unfortunately, it is impossible  in general. 
%Prior work showed that, unfortunately, it is impossible  in general. 

This paper charts a more complete landscape of the feasibility of perfect resilience. 
We first show a perhaps surprisingly large price of locality in static fast rerouting mechanisms: even when
source and destination remain connected by a linear number of link-disjoint paths after link failures, local rerouting algorithms cannot find any of them which leads to a disconnection on the routing level. 
This motivates us to study resilience in graphs which exclude certain dense minors, such as cliques or a complete bipartite graphs, and in particular, provide characterizations of the possibility of perfect resilience in different routing models.
%In particular, we provide characterizations of the possibility of perfect resilience in three main models: a model where routing rules can depend both on the source and the destination, a model where they can only depend on the destination, and a model where they do not depend on either of the two (but where packets need to visit all nodes of the network).
We provide further insights into the price of locality by showing impossibility results for few failures and investigate perfect resilience on Topology Zoo networks.
\end{abstract}

\section{Introduction}\label{sec:introduction}

Traditional communication networks can be modelled as distributed systems in which routers cooperate to compute efficient routes.
In particular, using protocols based on link state or distance vector algorithms, routers can---in a distributed manner---compute routing tables which induce shortest paths~\cite{chiesa2016traffic}.
These protocols can also naturally cope with failures: 
whenever one or multiple links fail, the distributed routing protocol is simply invoked again, triggered by the nodes incident to a failed link. 
The protocols are hence in some sense ``\emph{perfectly resilient}''~\cite{podc-ba}:
After reconvergence, the protocol re-establishes a path between any pair of nodes still physically connected, by dynamically updating their routing tables. 
Unfortunately, however, the recomputation and dynamic update of routing tables comes at the cost of slow reaction~time~\cite{frr-survey}.

Modern dependable communication networks hence additionally feature fully decentralized flow rerouting mechanisms which rely on \emph{static} routing tables and allow to react to link failures orders of magnitudes faster than traditional networks~\cite{frr-survey}.
Rather than invoking the distributed routing protocol when detecting a failure, these static fast rerouting mechanisms allow to predefine conditional failover rules at each router:
 these rules can depend only on \emph{local} information at a node $v$, and can hence be conditioned on the status of links incident to $v$ or the header of packets arriving at $v$, but not on failures in other parts of the network.
While this enables a very fast reaction, it raises the question of how such local rules can be defined to maintain a high resilience under multiple link failures.
Feigenbaum et al.~\cite{podc-ba} showed that achieving a perfect resilience using static fast rerouting
mechanisms is unfortunately impossible in general: 
the authors presented an example network in which it is not possible to predefine local failover rules which
ensure that as long as the underlying graph is connected, the routing tables induce a valid routing path to the destination.
In other words, there is a \emph{price of locality}: local fast rerouting comes at a cost of reduced resilience under multiple link~failures.

This paper provides a systematic analysis aiming to characterize the feasibility of perfect resilience using static fast rerouting, motivated by Feigenbaum et al.'s counterexample. 
Indeed, their work raises a number of interesting research questions, such as:

\begin{itemize}
	\item How significant is the price of locality? Is it at least possible to compute local failover rules which ensure connectivity on the routing level if the underlying network remains highly connected after the link failures? 
	
	\item How does the resilience depend on the model? 	
	What happens if we include the promise of high connectivity or few failures, respectively, if we aim for smaller routing tables and do not match on the packet source or not even on the destination---where are the boundaries between working algorithms and impossibility?
		This question is particularly interesting in the light of emerging software-defined networks which allow routers to match different header parts and thus implement different routing~models. 
\end{itemize}

\subsection{Contributions}

This paper aims to chart a more complete picture of the feasibility of perfect resilience with local fast rerouting, focusing on the most fundamental aspect: reachability.
We first show a perhaps surprisingly general negative result: 
%already an asymptotically negligible \todo{we fail nearly all links in our proof} number of link failures in a highly connected graph inevitably disconnects local failover routes in the network .
even when a large number of link-disjoint paths survive after link failures, local failover routes cannot leverage them to reach the destination.
Specifically, we prove the following price of locality:
even if we are promised that there remain $\Omega(n)$ disjoint paths between source and destination after failures (we refer to this scenario as \emph{$r$-tolerant} where $r=\Omega(n)$), it is impossible to pre-define static routing tables ahead of time which ensure connectivity without knowing these failures; here $n$ refers to the \# of nodes (\S\ref{sec:r-tolerance}). Prior work only showed impossibility for $1$-tolerance and left higher connectivity guarantees to future~work.

Motivated by this result, we study the feasibility of perfect resilience in graphs which exclude certain dense minors, such as cliques or a complete bipartite graphs.
We present an almost optimal characterization of resilience in the different models.
First, for a model in which routers can match both the source and the destination of a packet, we show that 
perfect resilience is impossible on \emph{any} graph which has a minor
$K_7$ or a minor $K_{4,4}$ which misses one link, 
but possible on $K_5$ and $K_{3,3}$ networks and their minors (\S\ref{sec:src}).\footnote{A $K_n$ is a complete graph with $n$ nodes, whereas a $K_{a,b}$ is a complete bipartite graph with $a$ respectively $b$ nodes in its two partitions.} 

In a model where routing rules can only match the packet destination, it is impossible to achieve perfect resilience on networks with minors $K_5$ and $K_{3,3}$ which miss one link; this characterization is
complete in the sense that we can show that perfect resilience is always possible
on $K_5$ and $K_{3,3}$ networks which only miss two links, and their minors (\S\ref{sec:routing-model}).

We also study the price of locality in scenarios in which the number of link failures is bounded (\S\ref{sec:few-removals}) as well as in scenarios in which the local routing rules do not even depend on the destination but where a packet needs to tour the entire network, i.e., visit all nodes under failures (rather than routing to a specific destination); we provide an exact characterization of perfect resilience in this model as well, touring is possible if and only if $G$ is outerplanar~(\S\ref{sec:touring}).
%

%Table~\ref{tbl:big1} and Figure~\ref{fig:scheme1} summarize our classification results.

Lastly, we also perform a small case study in \S\ref{sec:topo-case} on more than 250 Topology Zoo networks: around a third of all networks allow for perfect resilience in all models, while the classification of the remaining topologies depends on the routing model considered. For destination-based routing, our contributions allow us to to classify more than 30\% additional topologies than with previous results.

%Moreover, we show that all our algorithms use very small routing tables, which is an open question for failover routing in general, and show how to compose any set of failover algorithms for a construction related to block graphs.\klaus{maybe kick section out as it disrupts the story}

\subsection{Background and Related Work}\label{subsec:related-work}

The question of how to provide resilient routing in networks is a fundamental one and has been explored intensively in the literature already~\cite{frr-survey}. 
In particular, failover resiliency can impose a trade-off on, e.g., stretch or latency~\cite{ccr18failover,casa19,DBLP:conf/ancs/SchweigerF021}: ``\textit{a robust route is not necessarily the shortest route}''~\cite{DBLP:conf/ladc/SchroederD07}. 
Hence, it can be worthwhile to consider detours through highly connected components, in case further failures appear downstream~\cite{DBLP:conf/dsn/DuarteSC04}, and to such an end also investigate on how to rank the connectivity properties of nodes~\cite{DBLP:conf/complenet/CohenDS10}.
While such detours or failover routes can also be enhanced by shortcutting the paths before global convergence kicks in~\cite{DBLP:conf/ancs/ShuklaF21}, we in this paper focus on the aspect of resilience under rapid (instantaneous) reaction times.

Many existing approaches require dynamic routing tables~\cite{gafni-lr,corson1995distributed,DBLP:conf/spaa/BuschST03} which implies slow reaction times~\cite{frr-survey}, or the ability to rewrite or extend packet headers which introduces overheads and is not always possible~\cite{elhourani2014ip,infocom15,conext20}.
Our requirement of static failover tables and immutable headers also rules out the application of graph exploration techniques such as~\cite{hotsdn14failover,reingold,DBLP:journals/tcs/FoersterW16,DBLP:journals/tcs/MegowMS12} or the use of rotor routers~\cite{rrouter1,parallel-rotor,rotorrouter}.
Also classic routing algorithms for sensor networks, such as geographic routing~\cite{karp2000gpsr,kuhn2008algorithmic,kuhn2003worst,wattenhofer2004xtc,DBLP:conf/dagstuhl/Zollinger07}, require memory and are hence not applicable in our context. Furthermore, while there exist graph exploration algorithms which do not require any memory, e.g., for mazes consisting of a single wall (see e.g., the well-known right-hand rule~\cite{labyrinth}), these algorithms are transferrable only (if at all) to very simple graphs such as outerplanar graphs~\cite{apocs21resilience}.
Our model hence assumes an interesting new position in the problem space:
while it is not possible to use dynamic memory during routing (neither in the packet header nor in the routing table), it is possible to pre-process\footnote{Here we also refer to the SUPPORTED model~\cite{DBLP:conf/sigcomm/SchmidS13,DBLP:conf/infocom/FoersterH0S19,DBLP:journals/pomacs/FoersterKPR021}, which investigates on a fundamental level what can and what cannot be pre-processed.} conditional routing rules ahead of time, without knowing the actual failure~scenarios. 

The model considered in this paper was introduced by Feigenbaum et al.~\cite{podc-ba,DBLP:journals/corr/abs-1207-3732} and, in a slightly more restricted version, by Borokhovich et al.~\cite{opodis13shoot} in parallel work. 
While there has been interesting applied work on this problem, e.g.,~\cite{keep-fwd,plinko-full,schapira1,schapira2}, in the following, we will focus on related works providing theoretical insights.

\subsubsection{Ideal versus Perfect Resilience}
Several interesting results are due to Chiesa et al.\ who presented a technique which relies on a decomposition of the network into arc-disjoint arborescence covers~\cite{icalp16,robroute16infocom,DBLP:journals/ton/ChiesaNMGMSS17}:
any $k$-connected graph can be decomposed into a set of $k$ directed spanning trees~\cite{edmonds1973edge} (rooted at the same node, the destination) such that no pair of spanning trees shares a link in the same direction.
This allows to route packets along some arborescence until hitting a failure, after which the packet can be rerouted along a different arborescence.
This technique is particularly well-suited to provide a weaker notion of resilience, known as \emph{ideal resilience}~\cite{DBLP:journals/ton/ChiesaNMGMSS17}, which is defined for $k$-connected graphs (while the notion of perfect resilience applies to arbitrary graphs): given a $k$-connected network, static failover tables are called ideally resilient if they can tolerate any set of $k-1$ link failures.
In contrast, \emph{perfect resilience} is defined for all graphs: static failover tables are called perfectly resilient if they can tolerate any set of failures, as long as the destination is still connected to the packet's source after failures.
As thus perfect resilience is stronger than ideal resilience: perfect resilience implies ideal resilience, but not vice versa.
While Chiesa et al.'s paper already led to several follow up works~\cite{casa19,dsn19,srds19failover,ccr18failover,DBLP:conf/infocom/FoersterKP0T21}, it remains an open question whether ideal resilience can be achieved in general $k$-connected~graphs.

As mentioned above, already Feigenbaum et al.~\cite{podc-ba,DBLP:journals/corr/abs-1207-3732} proved that perfect resilience is impossible to achieve in general, by presenting a counterexample with 12 nodes. 
Foerster et al.~\cite{apocs21resilience} recently generalized this negative result by showing that it is impossible to achieve perfect resilience on any non-planar graph; furthermore, planarity is also not sufficient for perfect resilience. 
On the positive side,~\cite{apocs21resilience} showed that perfect resilience can always be achieved in outerplanar graphs, and also initiated 
the study of routing rules which can depend on the source.
In this paper, we significantly extend these results along several dimensions.

\subsection{Overview}

The remainder of this paper is organized as follows.
We introduce our formal model in \S\ref{sec:model}.
In \S\ref{sec:r-tolerance}, we show that maintaining connectivity with local failover rules is challenging already in highly connected graphs, and even if routing rules can depend on the source.
This motivates us to study perfect resilience on graphs with dense minors, in 
a model where routing tables can (\S\ref{sec:src}) or cannot (\S\ref{sec:dst}) depend on the source.
We then investigate the problem of perfect resilience under a bounded number of link failures (\S\ref{sec:few-removals}) and study a novel failover model, where routing cannot depend on source and destination but where a packet needs to visit the entire graph (\S\ref{sec:touring}).
In \S\ref{sec:topo-case} we then perform a case study on Topology Zoo networks to classify them w.r.t.\ perfect resilience.
We conclude our contribution and discuss future directions in~\S\ref{sec:conclusion}.
For better readability, some proof details and figures are deferred to the Appendix, beginning on page~\pageref{app-source}.

\section{Model}\label{sec:model}

We are given a communication network which we model as an undirected graph $G=(V,E)$, where the nodes represent \emph{routers} that are connected via \emph{links} $E$.
We define $n=|V|$, $m=|E|$, and write $V_G(v)$ and $E_G(v)$ for the neighbors and incident links of node $v$, respectively; if clear from the context, we will omit the subscript $G$. 
We will also write $V(G)$ and $E(G)$ for the nodes $V$, respectively links $E$, of a graph $G=(V,E)$. 
%We will denote the node degree by $d_v=|V_G(v)|$ (or simply $d$). %do we ever use this?
When talking about connectivity, we always refer to link connectivity, \emph{i.e.}, two nodes $v,w\ \in V(G)$ are $k$-connected if there are $k$ paths between $v$ and $w$ that do not share any links, such paths are also called link-disjoint paths.
The notations $K_n$ and $K_{a,b}$ refer to the complete graph with $n$ nodes, respectively the complete bipartite graph with $a$ and $b$ nodes in its partitions.
For the latter notations, when adding the superscript $-c$, i.e., $K^{-c}_n$ and $K^{-c}_{a,b}$, we remove $c$ links from the respective graphs.

The network is subject to link failures, which however are not known ahead of time, when the routers are configured.
We will refer to the set of links which will fail by $F\subset E$; failures are undirected.
The graph $G$ without links $F$ is denoted by $G \setminus F:= G(V,E\setminus F)$. 
Similarly, $G \setminus E'$ and $G \setminus V'$ denote the graph $G$ without the set of links in 
$E' \subset E$, respectively, the graph $G$ without the set of nodes $V' \subset V$ and their incident~links.

Each node $v\in V$ is configured with a local forwarding function $\pi(v)$, essentially a \emph{forwarding table}.
This forwarding table (or synonymously, \emph{routing table}\footnote{While forwarding table is the technically correct term, we will use the term interchangeably with the term routing table.}) is essentially a set of forwarding rules which include conditional failover rules that depend on the incident link failures.
 Specifically, the rules $\pi(v)$ of node $v$ can depend on (a subset of) the following~information:
\begin{itemize}
\setlength\itemsep{0em}
\item the set of incident failed links $F\cap E(v)$
\item the source $s$ of the to-be-forwarded packet at $v$
\item the destination $t$ of the to-be-forwarded packet at $v$
\item the incoming port (\emph{in-port}) from which the packet arrives at $v$
\end{itemize}

In this paper we aim to chart a landscape of resiliency results for different models, and we hence consider multiple combinations of the above information. 
However, all these models have in common that the routing table is pre-configured and static, and forwarding rules do not change the packet~header. 

A local routing algorithm is hence  simply a \emph{forwarding function} $\pi_v$ for each node $v$.
For example, in the most general model where all information can be accounted for, given a graph $G$ and a destination $t \in V(G)$, the function is 
$$\pi_v: ~~~ 2^{E(v)} \times V \times V \times E(v) \cup \{ \bot \} \mapsto E(v)$$ 
at each node $v \in V(G)$, where $\bot$ represents the empty in-port, i.e.\ the starting node of the packet. 
In other words, given the set of failed links $F \cap E(v)$ incident to a node $v$, the source and the destination, as well as the in-port, the forwarding function $\pi_v$ maps each incoming port (link) $e=(u,v)$ to the corresponding outgoing port (link).
We will call the union of the forwarding functions $\pi = (\pi_v)_{v \in V}$ the \emph{forwarding pattern}, or simply the \emph{routing}.
In the following, we will use the~notation
$$\pi_v^{s,t}(e,F) ~~~~ \text{     resp.   } ~~~~ \pi_v^{s,t}(u,F)  $$ 
to denote the link to which a packet arriving at $v$ via the link $e=(u,v)$ will be forwarded, given a failure set $F$ and in a model where the rule matches both source $s$ and destination $t$.
We will refer to these types of rules as \emph{source-destination-based routing}. 
Similarly, we will use the notation $\pi_v^{t}(e,F)$ resp.~$\pi_v^{t}(u,F)$ to denote the link to which a packet arriving at $v$ from a link $e=(u,v)$ will be forwarded, given a failure set $F$ and in a model where the rule matches only the destination $t$.
We refer to these types of rules as \emph{destination-based routing}. 

\begin{figure}[t]
  \begin{center}
%\resizebox{0.5\columnwidth}{!}{%
  \begin{tikzpicture}[shorten >=1pt]
  \tikzstyle{vertex}=[circle,fill=black!25,minimum size=17pt,inner sep=0pt]
  \tikzstyle{edge}=[thick,black,--]
	
	\node[vertex] (i) at (0,0) {$v$};
	\node[vertex] (1) at (1,1) {$v_1$};
	\node[vertex] (2) at (1,-1) {$v_2$};
	\node[vertex] (3) at (-1,-1) {$v_3$};
	\node[vertex] (4) at (-1,1) {$v_4$};
	
	\node[vertex] (ir) at (5+0,0) {$v$};
	\node[vertex] (1r) at (5+1,1) {$v_1$};
	\node[vertex] (2r) at (5+1,-1) {$v_2$};
	\node[vertex] (3r) at (5+-1,-1) {$v_3$};
	\node[vertex] (4r) at (5+-1,1) {$v_4$};
	
	\draw (i) to (1);
	\draw (i) to (2);
	\draw (i) to (3);
	\draw (i) to (4);
	\draw[dashed,thick,->, bend left,blue] (1) to (2);
	\draw[dashed,thick,->, bend left,blue] (2) to (3);
	\draw[dashed,thick,->, bend left,blue] (3) to (4);
	\draw[dashed,thick,->, bend left,blue] (4) to (1);
	
	\draw (ir) to (1r);
	\draw (ir) to (2r);
	\draw (ir) to (3r);
	\draw (ir) to (4r);
	
	\draw[dashed,thick,->, bend left,red] (1r) to (2r);
	\draw[dashed,thick,->, bend left,red] (2r) to (3r);
	\draw[dashed,thick,->, bend left,red] (3r) to (2r);
	\draw[dashed,thick,->, bend right,red] (4r) to (3r);
	
\end{tikzpicture}
%} %end resizebox
  \end{center}
%\vspace{-5mm}
\caption{On the left is an example for a forwarding pattern for the node $v$ that follows a cyclic permutation $(v_1,v_2,v_3,v_4)$: packets coming from $v_1$ are forwarded to $v_2$, packets from $v_2$ to $v_3$, from $v_3$ to $v_4$, and from $v_4$ to $v_1$. Other examples for cyclic permutations for the node $v$ would be, e.g., $(v_1,v_3,v_4,v_2)$ or $(v_1,v_4,v_3,v_2)$. On the right is an example for a forwarding pattern for the node $v$ that does not follow a cyclic permutation, as, e.g., $v$ will not route any incoming packet to $v_1$ or $v_4$.}
	\label{fig:cyclic-explanation}
\end{figure}
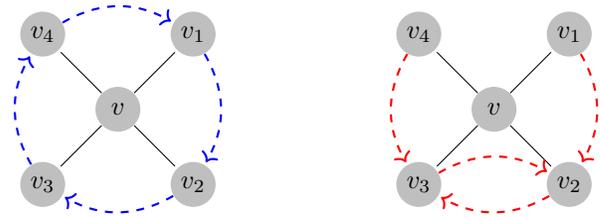

Note that we do not require these forwarding patterns to follow some sort of cyclic permutation (as in, e.g., Figure~\ref{fig:cyclic-explanation}) of the out-ports for neither of the routing flavours .

%\review{ In particular, the concept of routing "in a cyclic permutation" is used throughout the paper, without being clearly defined (maybe it's standard, and I could figure it out from context, but a single sentence clarification would make a big difference).}

We will call a forwarding pattern $\pi$ \emph{$r$-resilient} if for all $G$ and all $F$, where $|F| \leq r$, the forwarding pattern routes the packet from all $v \in V$ to any destination $t$ when $v$ and $t$ are connected in $G \setminus F$. 
Note that the restriction that source and destination must remain connected when removing the links in the failure set $F$ implies that the connectivity of the graph does not play a big role. E.g., consider a graph $G$ which consists of a cliques of size $l$ an one extra node connected to the clique with one link. While the connectivity of $G$ is one, it is easy to construct forwarding patterns that tolerate two failures for packets originating from the extra node if the remaining graph stays connected.  
A forwarding pattern is \emph{perfectly resilient} if it is $\infty$-resilient: the forwarding always succeeds in the connected component of the destination, for all destinations.
Let $A_p(G,s,t)$ be the set of such perfectly resilient patterns (algorithms), respectively $A_p(G,t)$, $A_p(G)$ for the different models depending only on the destination or not even that; we abbreviate these versions by $A_p$ when the context is~clear.

To explore the achievable resilience of local fast rerouting algorithms beyond perfect resilience, we in this paper are also interested in a relaxed notion of resiliency, where we are given the promise of high connectivity after failures:

\begin{definition}[$r$-tolerant]\label{def:r-tolerant}
A forwarding pattern $\pi^{s,t}$ is called $r$-tolerant on a graph $G$, if it can guarantee reaching the destination $t$ from source $s$ under the assumption that $s$ and $t$ remain $r$-connected under failures.
\end{definition}

Observe that $r=1$-tolerance corresponds to perfect resilience and that, for $r<r'$, if we have $r$-tolerance, we also obtain $r'$-tolerance: the failure sets $\mathcal F_r$ that retain $r$-connectivity are a superset of the failure sets $\mathcal F_{r'}$ that retain $r'$-connectivity. For example, a perfectly resilient algorithm ($1$-tolerant) is also $2$-tolerant. 

\section{On the Price of Locality}\label{sec:r-tolerance}

Before studying perfect resilience in detail, we first consider a weaker notion
of resilience: the design of local rerouting functions for scenarios where the connectivity remains larger than one after failures.
We derive a surprisingly strong negative
result on what can be achieved with static fast rerouting:

It is generally \emph{impossible} to be
$\Omega(n)$-tolerant, even when forwarding rules can depend on both
source and destination. 

Prior work just showed that $1$-tolerance is impossible in general, but the details for higher connectivity promises were left unanswered.
Indeed, at first it seems that if we are guaranteed that a linear number of paths exist after failures between source and destination, then surely fast failover mechanisms should be able to leverage this high connectivity.
However, we show next that this intuition is false.

\subsection{Intuition and Example}
\begin{figure}[t]
  \begin{center}
%\resizebox{0.65\columnwidth}{!}{%
  \begin{tikzpicture}[shorten >=1pt]
  \tikzstyle{vertex}=[circle,fill=black!25,minimum size=17pt,inner sep=0pt]
  \tikzstyle{edge}=[thick,black,--]
	\tikzset{snake it/.style={decorate, decoration=snake}}
	
	\draw [dotted] (0,0) ellipse (1.4cm and 1.7cm);
	
	\draw [dotted] (3,0) ellipse (1.4cm and 1.7cm);
	
	\node[vertex] (s) at (-1,0) {$s$};
	\node[vertex] (1) at (0.8,0.8) {$v_1$};
	\node[vertex] (2) at (0.8,-0.8) {$v_3$};
	\node[vertex] (11) at (2.2,0.8) {$v_1'$};
	\node[vertex] (21) at (2.2,-0.8) {$v_3'$};
	\node[vertex] (t) at (4,0) {$t$};
	
	\node[vertex] (fl) at (0.8,0) {$v_2$};
	\node[vertex] (fr) at (2.2,0) {$v_2'$};
%
	%%\node[vertex] (1) at (0,0) {$v_1$};
%
	\draw (1) to (11);	
	\draw[dashed,red,thick] (2) to (21);	
	\draw[snake it] (s) to (1);
	\draw[snake it] (s) to (2);
	\draw[snake it] (t) to (11);
	\draw[snake it] (t) to (21);
	\draw[snake it] (t) to (fr);
	\draw[snake it] (s) to (fl);
	
	\draw (fl) to (fr);
	
	\node (bot) at (1.5,1.8) {};
	\node (top) at (1.5,-1.9) {};
	\draw[dotted, blue,thick] (bot) to (top);
	
	%\node (f1l) at (0.8,0.4) {};
	%\node (f1r) at (2.2,0.4) {};
	%\node (f2l) at (0.8,0.0) {};
	%\node (f2r) at (2.2,0.0) {};
	%\node (f3l) at (0.8,-.4) {};
	%\node (f3r) at (2.2,-.4) {};
	
	%\draw[dashed,red] (f1l) to (f1r);
	%\draw[dashed,red] (f2l) to (f2r);
	%\draw[dashed,red] (f3l) to (f3r);

\end{tikzpicture}
%}
  \end{center}
	%\vspace{-6mm}
\caption{After failures (in dashed red), $s$ and $t$ remain 2-connected, as there are two crossings across the blue dotted cut. Assume $v_1$ does not forward to $v_1'$ and $v_2$ does not forward to $v_2'$. Then, as the link  between $v_3$ and $v_3'$ has failed, it is impossible to reach $t$ from~$s$. Note that from a local point of view, $v_1,v_2,v_3$ are unaware of the failures at each other, and hence to guarantee $2$-tolerance, at least two of $v_1,v_2,v_3$ must forward across the blue dotted cut, to their respective $v_1',v_2',v_3'$, if still possible after failures.}
	\label{fig:rimp}
	%\vspace{-3mm}
\end{figure}
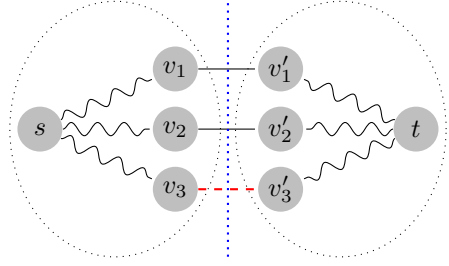

Intuitively, the more highly connected the topology is after failures, the easier it 
should be to ensure connectivity also with local static rerouting. 
However, as we will illustrate here on complete networks, this additional topological connectivity is only marginally useful.
Concretely, while an $r$-tolerant algorithm in principle has more flexibility, 
in the sense that it can afford to \emph{not} explore a certain route at all (as there are for sure alternative routes), and hence e.g., avoid potential loops, this additional connectivity is hard to exploit \emph{locally}: such a choice can only be made $r-1$ times for an $r$-tolerant algorithm, among \emph{all nodes} in the graph. In other words, the flexibility is restricted globally, while decision making is inherently local. 
We refer to Fig.~\ref{fig:rimp} for an illustration.
%
%As a result, a forwarding function that does not explore a link can only be assigned to few nodes. \juho{I edited this sentence, but it still feel like repetition of the previous sentences.}

%\begin{figure}[htbp]
	%\centering
		%\includegraphics[width=0.45\textwidth]{figs/silver-bullet.JPG}
	%\caption{silver-bullet for t=2}
	%\label{fig:silver-bullet}
%\end{figure}

\subsection{Impossibility of $r$-Tolerance in General}\label{subsec:r-tol1}

We show the following general impossibility result, namely that $r$-tolerance is impossible in general, already on instances that only grow linearly with $r$:

\begin{theorem}\label{thm:r-tolerant-impossible}
Let $r \in \mathbb{N}$. The complete graph with $3+5r$ nodes does not allow for an $r$-tolerant  forwarding pattern $\pi^{s,t}$.
\end{theorem}

\begin{IEEEproof}
From $K_{3+5r}$ we choose 5 nodes $V_5 =\{v_1,v_2,v_3,v_4,v_5\}$ not including $s$ and $t$.
Consider all triples $a,b,c \in V_5$, such that, if $b$ has a degree $2$ after failures (connected to only $a,c$), then $b$ will not forward a packet from $a$ to $c$. 
If such a triple exists, then leave the path $s-a-b-c-t$ intact after failures and remove all other links of $a$,$b$, and $c$.  We have constructed a partial failure set and a path from source to target that is not used by the forwarding pattern under this partial failure set.

If such a triple does not exist, then all nodes in $V_5$, with degree 2 after failures, will route in a permutation, assuming their neighbors are from $V_5$.
Without loss of generality (W.l.o.g.), leave the path $s-v_1-v_2$ intact after failures, but fail all other links of $v_1$. Then, for $v_2$, leave only the links to $v_3,v_4,v_5$ alive after failures. 
If the routing of $v_2$, coming from $v_1$, does not enter a permutation containing all $4$ neighbors, with $v_i \neq v_1$ missing, then we fail all links incident to $v_3$, $v_4$, and $v_5$, except the links to $v_2$ and the link between $v_i$ and $t$. Now, the packet coming from $v_1$ to $v_2$ will not reach $v_i$, and hence we lose one path to destination. On the other hand, if $v_i = v_1$ is missing, then we fail all links incident to $v_3$, $v_4$, and $v_5$ and the packet is trapped in the 5-node construction without returning to $s$ via $v_1$.

Else the routing is a cyclic permutation. Assume w.l.o.g.\ the cyclic ordering for $v_2$ is $v_1,v_3,v_4,v_5$. We then fail all links incident to $v_3$, $v_4$, and $v_5$, except the link to $v_2$ and the links $(v_4,t)$ and $(v_2,v_5)$. 
Now, the packet is routed $s-v_1 - v_2 - v_3 - v_5 - v_2$. The packet will then go to $v_1$ and start a loop --	we lose one path to the destination, namely via $v_4$.
We repeat the construction above $r$ times in total, always picking a new set of 5 nodes. 
Each time we either lose one path to the destination or find a routing loop.
If we lose $r$ paths, then the construction is complete, but we also need to consider the case where we are trapped in a routing loop in a 5-node gadget, as then the $st$-connectivity is $r-1$.
To this end,
we use the one remaining node $v$ from $K_{3+5r}$, and leave it connected to $s$, but fail all its other incident links except $(v,t)$.
W.l.o.g., we can assume that $v$ is last in the visiting order of $s$.
If we lose $r$ paths, then we disconnect $v$ from $t$ and hence the packet loops permanently, as none of the other 5 node constructions allow passing to the destination.
If we do not lose $r$ paths, then the path $s-v-t$ restores $st$-connectivity to $r$, as promised, but the packet loops in one of the 5 node constructions.
\end{IEEEproof}

Note that $r$-tolerance is preserved under iteratively taking subgraphs, \emph{i.e.}, if $G$ allows for $r$-tolerance, then every $G'\subset G$ allows for $r$-tolerance as well.
The reason is that we can obtain $G'$ as a component of $G$ by failing the missing links.

\begin{corollary}
Let $r \in \mathbb{N}$. If a graph $G$ has $K_{3+5r}$ as a subgraph, then $G$ has no $r$-tolerant  forwarding pattern $\pi^{s,t}$.
\end{corollary}

\subsection{$r$-Tolerance and Minors}\label{subsec:r-tol2}

Even though $r$-tolerance is preserved under taking subgraphs, we next show that $r$-tolerance is not preserved for graph minors, for all $r \geq 2$. This is in contrast to the result of Foerster et al.~\cite{apocs21resilience}, who showed that it is preserved for $r=1$.
In other words, there is a fundamental distinction between $r=1$ and all larger $r$, which is to be investigated in future work:

\begin{theorem}\label{thm:no-minor-r}
For each $r \in \mathbb{N} \colon r \geq 2$ holds: There exists an $r$-tolerant graph $G$ with a minor $G'$ that is not $r$-tolerant.
\end{theorem}

\begin{IEEEproof}
Given parameter $r>1$, let the construction from Theorem~\ref{thm:r-tolerant-impossible} be denoted by $G'$.
We will next show how to build a graph $G$, s.t.\ $G$ is $r$-tolerant and $G'$ is a minor of $G$:
Given $G'$, add a new source node $s'$, connect it with $r-1$ paths to $s$, and add the link $(s',t)$. 
An algorithm that is $r$-tolerant on this new graph simply routes from $s'$ to $t$ via the direct link; if that link fails, the $r$-tolerance promise does not hold. 
Now, observe that the graph construction from Theorem~\ref{thm:r-tolerant-impossible} is a minor of the above construction (obtained by merging $s',s$, as well as the paths between them, and removing the link  between $s'$ and $t$), i.e., the existence of an $r$-tolerant forwarding pattern does not imply the existence of an $r$-tolerant forwarding pattern for minors, for any $r \geq 2$.
\end{IEEEproof}

We note that if $s$ and $t$ are less than $r$-connected before the failures occur, $r$-tolerance trivially holds: $r$-tolerance is a promise problem, only considered under high connectivity.
On complete graphs, $r$-tolerance is also trivial for $K_{r+1}$, as a removal of the source-destination link removes the promise of $r$-connectivity.
We can slightly extend this result and give promises for connectivity beyond $r$:

\begin{theorem}\label{thm:r-dist-2}
For each $r \in \mathbb{N}$ $K_{2r+1}$ admits $r$-tolerance.
\end{theorem}
\begin{IEEEproof}
Foerster et al.~\cite[Theorem 6.1]{apocs21resilience} showed that perfect resilience can be maintained if source and destination have distance at most two after failures, and we now leverage their algorithm.
Assume that the link between source and destination fails on $K_{2r+1}$ -- else the statement holds by routing in a single direct hop.
When source $s$ and destination $t$ on $K_{2r+1}$ remain $r$-connected, then $s$ is connected to at least $r$ neighbors $V_1$ different from $t$ and $t$ is connected to at least $r$ neighbors $V_2$ different from $s$.
Besides source and destination, $K_{2r+1}$ has only $2r-1$ nodes, and hence $|V_1 \cap V_2| \geq 1$, \emph{i.e.}, a path of length 2 exists between source $s$ and destination $t$.
\end{IEEEproof}

We next briefly investigate complete bipartite graphs.
If source and destination are in the same part, then the distance-2 algorithm~\cite[Theorem 6.1]{apocs21resilience} applies if the other part has at most $2r-1$ nodes, as in the proof above.  
If source and destination are in different parts, then the distance-2 algorithm is no longer directly applicable, as every route besides the direct source-destination link has a length of at least 3.
However, we can extend the distance-2 forwarding pattern to distance 3 in complete bipartite graphs, as described in the following proof:

\begin{theorem}\label{thm:algo-dist-3-bipartite}
For all bipartite graphs $G$ there is a forwarding pattern $\pi^{s,t}$ that 
can guarantee reaching the destination $t$ from source $s$, if $s$ and $t$
%succeeds if source $s$ and destination $t$ 
are at distance at most 3 in $G\setminus F$.
\end{theorem}
\begin{IEEEproof}
 First, whenever the destination is a neighbor, we route to it, as highest priority. Else, the source and each neighbor of the source routes in a cyclic permutation. If a node is not the source or a neighbor of the source, then the packet bounces back (distance to source $=2$).
We only visit a node $v$ of distance 3 if $v=t$. Moreover, if the $st$-distance is at most $3$, we will also reach $t$ from $s$, as (without the hop to the destination or stopping when finding the destination), our algorithm traverses all links $E_1$ incident to the destination and all links $E_2$ adjacent to those in $E_1$, \emph{i.e.}, $t$ is found with a distance of at most $2$, and if the distance is exactly $3$, the last link is adjacent to a link in $E_2$.
\end{IEEEproof}

Applying Theorem~\ref{thm:algo-dist-3-bipartite} to complete bipartite graph yields:

\begin{theorem}\label{thm:r-dist-3}
For each $r \in \mathbb{N}$ $K_{2r-1,2r-1}$ admits $r$-tolerance.
\end{theorem}

% \begin{IEEEproof}
% We now perform a case distinction whether $s,t$ are in the same part of the partition, w.l.o.g.\ the first part.
% We start with the case where this is true. %second partition at most 2r-1 nodes!
% Again, due to $r$-connectivity after failures, the source $s$ retains at least $r$ neighbors $V_1$ from the second part, and the destination $t$ retains at least $r$ neighbors $V_2$ from the second part as well.
% Hence, $|V_1 \cap V_2| \geq 1$ due to each part having at most $2r-1$ nodes, \emph{i.e.}, a route of length 2 exists between $s$ and $t$.

% We next consider the remaining case where w.l.o.g.\ the destination is in the second part.
% If the link $(s,t)$ exists we are done immediately and hence assume it has failed.
% Else, $s$ has at least $r$ neighbors $V_1$ in the second part and $t$ has at least $r$ neighbors $V_2$ in the first part, with $s$ and $t$ not being in $V_1 \cup V_2$.
% As without $s$ and $t$ each part has only a size of $2r-2$, and due to $r$-connectivity, there must be at least $r$ links between the first and second part, not counting those incident to $s$ and $t$. 
% Hence, due to $r$-tolerance, there is at least 1 link $(u,w)$ alive between $V_1$ and $V_2$, \emph{i.e.}, source and destination have a distance of at most $3$ via the path $s-u-w-t$.
% \end{IEEEproof}

\begin{IEEEproof}
  Let $A$ and $B$ be the two parts of the bipartite graph $K_{2r-1,2r-1}$.
Assume w.l.o.g\ $s\in A$. 
  We now perform a case distinction whether $t\in A$.
We start with the case where this is true. %second partition at most 2r-1 nodes!
Due to $r$-connectivity after failures, the source $s$ retains at
least $r$ neighbors $V_s$ in $B$, and the destination $t$ retains at
least $r$ neighbors $V_t$ in $B$ as well.
Hence, $|V_s \cap V_t| \geq 1$ due to $B$ having at most $2r-1$ nodes, \emph{i.e.}, a route of length 2 exists between $s$ and $t$.

We next consider the remaining case where w.l.o.g.\ the destination is
in $B$ the second part.
If the link $(s,t)$ exists we are done immediately and hence assume it has failed.
Else, $s$ has at least $r$ neighbors $V_s$ in $B$ and $t$ has at least $r$ neighbors $V_t$ in $A$.
Pick $u\in V_s$: $u$ has at least $r-1$ neighbors $V_u$ in $A-\{s\}$. Since
$|A-\{s\}|=2r-2$, $V_t\cap (A-\{s\})$ (of size $r$) and $V_u \cap
(A-\{s\})$ (of size $r-1$) necessarily intersect. Let $w$ a node in
this intersection: $w$ is neighbor of both $u$ and $t$, hence source and destination have a distance of at most $3$ via the path $s-u-w-t$.
\end{IEEEproof}

We recall that $r$-tolerance is preserved for all subgraphs and obtain the following corollary:

\begin{corollary}\label{corr:r-tolerance-positive}
For each $r \in \mathbb{N}$ it holds that $K_{2r+1}$ and $K_{2r-1,2r-1}$ and all their respective subgraphs admit $r$-tolerance.
\end{corollary}

\section{Perfect Resilience with Source}
\label{sec:src}

Given our insights on the feasibility of local fast rerouting in more highly connected graphs, we now turn to studying perfect resilience: resilience in scenarios where arbitrary links can fail, as long as the graph remains connected.
Recall that we aim to chart a landscape of perfect resilience in this paper, and in this section, we start analyzing a model where routing rules can depend both on the source and the destination of a packet. In the next section, we will then consider the scenario where rules can only depend on the destination. 
%In this section we hence provide a characterization of the impossibility 
%and possibility of perfect resilience with source information. 
Given the result of the previous section, our characterization will revolve around graphs which feature dense minors before failures occur. 

\subsection{Impossibility Results}\label{subsec:source-negative}

%While it thus is possible to achieve perfect resilience on five nodes, no algorithm for $K_7$ exists. 
We first show that it is impossible to achieve perfect resilience on complete graphs with seven nodes.
%
%The main idea of the proof is shown in Fig.~\ref{fig:k7-imposs}, where we illustrate the situation after failures.
%
%The complete proof of \Cref{thm:nok7-1} is deferred to \Cref{app-source} due to space constraints, along with the other proofs of this section.

\begin{theorem}\label{thm:nok7-1}
The complete graph with seven nodes, minus one link, does not allow perfect resilience, i.e., $A_p(K_7^{-1},s,t)=\emptyset$.
\end{theorem}

Note that when considering perfect resilience, it is at most as hard to route in a subgraph as one can treat the missing edges as failed edges to simulate a forwarding pattern of a supergraph. We will see in our case study in Section~\ref{sec:topo-case} that removing one link makes a difference for the applicability of our results.

\begin{IEEEproof}[Proof Sketch]
The proof idea is depicted in Fig.~\ref{fig:k7-imposs-1}: as any of the neighbors of $v_2$ could be the only way to reach $t$, $v_2$ must route in a cyclic permutation if no incident links fail, analogously for the neighbors of $v_2$ if they have a degree of two and do not neighbor $s,t$ after failures.
Hence, for every permutation chosen for $v_2$, the failures of the surrounding nodes can be adjusted such that a routing loop occurs.
\begin{figure}[b]
%\vspace{-3mm}
  \begin{center}
\resizebox{0.55\columnwidth}{!}{%
  \begin{tikzpicture}[shorten >=1pt]
  \tikzstyle{vertex}=[circle,fill=black!25,minimum size=17pt,inner sep=0pt]
  \tikzstyle{edge}=[thick,black,--]

	\node[vertex] (s) at (-1.25,-1) {$s$};
	\node[vertex] (5) at (0,0) {$v_5$};
	\node[vertex] (1) at (0,-2) {$v_1$};
	\node[vertex] (2) at (1.25,-1) {$v_2$};
	\node[vertex] (3) at (2.5,0) {$v_3$};
	\node[vertex] (4) at (2.5,-2) {$v_4$};
	\node[vertex] (t) at (3.75,-1) {$t$};
	
	\draw (s) to (1);
	\draw (1) to (2);
	\draw (5) to (2);
	\draw (3) to (2);
	\draw (4) to (2);
	\draw (t) to (4);
	\draw (5) to (3);
\end{tikzpicture}
}
  \end{center}
	%\vspace{-3mm}
\caption{$K_7$ (without $s$-$t$ link) impossibility, when $v_2$ routes with the permutation $v_1,v_3,v_4,v_5,v_1$. As $v_3,v_5$ also route in a cyclic permutation, due to local indistinguishability, packets loops permanently in $v_2-v_3-v_5-v_2$.}
	\label{fig:k7-imposs-1}
%\todo{$K_7$ has 21 link s, here only 7 remain and 14 are gone, but maybe APOCS gives better number..}
\end{figure}
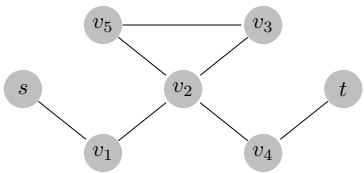
\end{IEEEproof}

%We briefly also investigate the maximum number of links that need to be removed to achieve impossibility on the~$K_7$.
%
The above proof never removes more than 15 links, hence:

\begin{corollary}\label{corr:edge-removal-k7}
Even under the promise that at most 15 links fail and there is an $st$-path, the complete graph $K_7$ with seven nodes does not allow for a forwarding pattern $\pi^{s,t}$ that 
%always succeeds
can guarantee reaching the destination $t$ from source $s$ 
 if $|F|\leq 15$.
\end{corollary}

%We next also show impossibility on
%provide a different construction for impossibility on 
 The impossibility can be shown on the $K_{4,4}$ analogously, however, as it is much sparser than the $K_7$, we need to remove fewer links. 
We refer to the Appendix for details.
%\review{Please revise the following sentence: "We omit the case distinction details."}
%We omit the case distinction details.
%
%Here, we cannot directly apply Corollary~\ref{corr:s-orbit-ext}, but rather need to carefully handcraft further permutation routing arguments.

\begin{theorem}\label{corr:nok44_with_source-1}
The complete bipartite graph with eight nodes, four in each part, minus one link, does not allow for perfect resiliency, i.e., it holds that  $A_p(K_{4,4}^{-1},s,t)=\emptyset$.
\end{theorem}

%\klaus{todo proof missing}

%We again investigate the maximum number of links that need to be removed to achieve impossibility on the~$K_{4,4}$.

%We again briefly investigate the number of link  failures in the above proof.
%Here we constructed the failure sets manually for each argument and did not leverage Corollary~\ref{corr:s-orbit-ext}, using at most $11$ link  failures.

\begin{corollary}\label{corr:edge-removal-k44}
Even under the promise that at most 11 link fail and that there is an $st$-path, the complete bipartite graph $K_{4,4}$ does not allow for a forwarding pattern $\pi^{s,t}$ that 
can guarantee reaching the destination $t$ from source $s$ 
%always succeeds
if $|F|\leq 11$.
\end{corollary}

\subsubsection{Generalization of Impossibility: Minor Relationships}
It was previously shown that if a graph $G$ allows for perfect resiliency with the source, so do all minors of $G$~\cite[\S4]{apocs21resilience}. 
Hence and in particular, all graphs containing a $K_{4,4}$ or a $K_{7}$ minus one link as a minor do not allow for perfect resilience.

\subsection{Possibility Results}\label{subsec:source-positive}

We now provide positive results on when perfect resilience is achievable.
Interestingly, as we will see, we can almost perfectly complement above impossibility results, by  providing algorithms for graphs characterized by less dense minors. 
We start by giving an algorithm for the $K_5$ and its subgraphs:
%The complete proofs are deferred to \Cref{app-source} due to space constraints.

%\stefan{fixme: from here}

%Algorithm~\ref{alg:K5_with_source} contains a more detailed pseudo code description.

\floatname{algorithm}{Algorithm}
\begin{algorithm}[t]
	\caption{ Perfectly resilient algorithm for $K_5$ and its minors} \label{alg:K5_with_source}
	\begin{algorithmic}[1]

	\Require packet from source $s$ to be delivered to destination $t$, local failure set $F_i$, identifiers $u < v < w$
	\Ensure forwarding port decision at node $i$

	\If{$(i,t) \notin F_i$}
		\State \textbf{send} to $t$
	\ElsIf{$i = s$}
		\If{exactly one neighbor $v$ is reachable}
			\State send to $v$
		\ElsIf{exactly two neighbors $u, v$ are reachable, $u < v$}
			\State \textbf{if} $inport=\bot$ \textbf{then} send to $u$
			\State \textbf{else} send to $v$ \Comment{ignore inport}
		\ElsIf{exactly three neighbors $u, v, w$ are reachable, $u < v < w$}
			\State \textbf{if} $inport=\bot$ \textbf{then} send to $u$
			\State \textbf{else if} $inport=w$ \textbf{then} send to $v$
			\State \textbf{else} \textbf{then} send to $w$ \Comment{coming from $u$ or $v$}
		\EndIf
	\Else
		\State \textbf{if} $inport=s$ \textbf{then} send to the neighbor with lowest ID (not $s$) or send back to $s$ if no other choice
		%\State \textbf{else if} there is a reachable neighbor $u$ and $inport \neq u$ \textbf{then} send to $u$
		\State \textbf{else if} there is a reachable neighbor $x$ and $inport \neq x$ \textbf{then} send to $x$
		\State \textbf{else if} $s$ is reachable \textbf{then} send to $s$
		\State \textbf{else } send back to $inport$
	\EndIf
	\end{algorithmic}
\end{algorithm}

\begin{theorem}\label{thm:k5_with_source}
For all graphs with at most five nodes Algorithm~\ref{alg:K5_with_source} describes a forwarding pattern matching on the source guaranteeing perfect resilience.
\end{theorem}

\begin{IEEEproof}
  We proceed by showing that packets routed with Algorithm~\ref{alg:K5_with_source} reach the destination for all possible distances between source and destination after failures.
  
  By showing it for $K_5$, we directly show correctness for all minors of $K_5$ as well, due to~\cite[Corollary 4.2]{apocs21resilience}.

  If the distance between source and destination is one, Line 2 of the algorithm ensures the packet arrives at its destination directly.

  If the distance is two, there are four non-isomorphic candidate graphs on which  a packet could visit all other nodes before visiting $t$, $G_1, G_2, G_3, G_4$ with $V=\{s,t,x,y,z\}$ 
  and link  sets $E_1=\{(x,y), (y,s), (s, z), (z,t)\}$,  $E_2=E_1 \cup \{(x,s)\}$, $E_3=\{(x,s), (s,y), (y,z), (z,t), (s,z)\}$ and $E_4=\{(s,x), (s,y), (s,z), (z,t))\}$, after removing the failed links respectively. 
  Depending on how we order the IDs for $x,y,z$ for $E_1$, the algorithm may first explore $x$ before returning to $s$ but it will definitely visit $z$ via $y$ and thus find $t$.
  For $E_2$, the algorithm will head straight towards $t$ if $z$ has the lowest identifier. If $y $ is the lowest identifier, the algorithm will visit the nodes in the order $s, y, x, s,z,t$ regardless of the order of the identifiers of $x,y$. 
  For $E_3$, the sequence of nodes visited starts with $s, x, s$ if $x$ has the lowest identifier, followed by $y,z,t$ if $y=v$ and $z=w$ or $z,t$ otherwise. If $y$ has the lowest identifier the sequence is $s,y,z,t$, if $z=y$ it is $s,z,t$. 
  For $E_4$ the algorithm guarantees that all neighbors of the source are visited if the previous ones did not connect to the destination as the nodes will send the message back if they cannot forward it to $t$.
  Note that for subgraphs of $G_1, G_2, G_3, G_4$ where $(s,x)$ is missing and/or $(s,y)$ is missing from $G_4$ the destination is reached in at most the same number of steps as well by the same line of arguments, as some detours will not be taken.

  If the distance is three, six non-isomorphic candidate graphs exist where a packet could visit all other nodes before visiting $t$,  $G'_1, G'_2, G'_3, G'_4, G'_5, G'_6$ with $V=\{s,t,x,y,z\}$ and link  sets $E'_1=\{(x,s), (s,y), (y,z), (z,t)\}$,  $E'_2=E_1 \cup \{(x,y)\}$, $E'_3=\{(s,x), (x,y), (y,t), (z,y)\}$, $E'_4=\{(s,x), (x,y), (y,t), (z,x))\}$, $E'_5=E_4\cup\{(z,y)\}$, and $E'_6=E_5\cup\{(z,t)\}$, after removing failed link respectively.
  For $E'_1$ the algorithm will forward packets on its direct path to the destination if $y=u$. Otherwise there might be a detour to $x$ first. 
  For $E'_2$,  if $x=u$ then the sequence of nodes visited is $s,x,y,z,t$, if $y=u, x=v$ then it is $s,y,x,s,x,y,z,t$, if $y=u, z=v$ or $z=u, y=v$ then no detour is taken and it the remaining case with $z=u, x=v$ the path used Is $s, x, y,z,t$.
  For $E'_3$, the path taken is $s,x,y,t$ and for $E_4$ a visit to $z$ might be included but no loop introduced.
  For $E'_5$, $z$ is visited if $z<y$ leading to a path of $s, x, z, y,t$ and $s,x,y,t$ otherwise. 
  In the last graph $E'_6$, visiting $z$ would lead to a shortcut to $t$ and in both cases $t$ is reached.
  Note that for subgraphs of $G'_1$ without $(s,x) $ and $G'_3, G'_4$ without link to $z$ the destination is reached in at most the same number of steps as well by the same line of arguments, as some detours will not be taken.

  If the distance is four, the nodes form a chain and the algorithm ensures that all nodes forward the packet until it reaches it destination.
\end{IEEEproof}

\noindent We obtain further positive results for complete bipartite graphs and refer to the Appendix for proof details.

\begin{theorem}\label{thm:k33_with_source}
There exists a forwarding pattern matching on the source and guaranteeing perfect resilience for the complete bipartite graph with three nodes in each part, and its minors.
\end{theorem}

\section{Perfect Resilience without Source}\label{sec:routing-model}
\label{sec:dst}

%\vspace{1mm}
%\resizebox{\columnwidth}{!}{%
\begin{figure*}
\begin{minipage}[b]{0.70\linewidth}
\resizebox{\columnwidth}{!}{%
\noindent\begin{tabular}{@{}ll@{\hspace{0.5cm}}ll@{\hspace{0.5cm}}ll@{\hspace{0.5cm}}}
@ $v_1$ & $\perp: v_2,v_3,v_4$ & $v_3: v_2,v_4,v_3$ & $v_4: v_2,v_3,v_4$ & ($v_2$: when we visit both we are done) \\
@ $v_2$ & $\perp: v_1,v_3,v_4$ & $v_3: v_1,v_4,v_3$ & $v_4: v_1,v_3,v_4$ & ($v_1$: when we visit both we are done) \\
@ $v_3$ & $\perp: v_2,v_1,v_4$ & $v_1: v_2,v_4,v_1$ & $v_2: v_1,v_4,v_2$ & $v_4: v_1,v_2,v_4$ & \\
@ $v_4$ & $\perp: v_1,v_2,v_4$ & $v_1: v_2,v_3,v_1$ & $v_2: v_1,v_3,v_2$ & $v_3: v_2,v_1,v_3$ & \\
\end{tabular}
}
\vspace{2.5mm}
\caption{Routing table to visit both neighbors of $t$ in Fig.~\ref{fig:k5-2-poss} under perfect resilience.}\label{fig:tab:k52}
\end{minipage}
\begin{minipage}[b]{0.29\linewidth}
  \begin{center}
\resizebox{0.5\columnwidth}{!}{%

  \begin{tikzpicture}[shorten >=1pt]
  \tikzstyle{vertex}=[circle,fill=black!25,minimum size=17pt,inner sep=0pt]
  \tikzstyle{edge}=[thick,black,--]

	\node[vertex] (1) at (0,0) {$v_1$};
	\node[vertex] (2) at (0,-2) {$v_2$};
	\node[vertex] (3) at (2,0) {$v_3$};
	\node[vertex] (4) at (2,-2) {$v_4$};
	\node[vertex] (t) at (-1.25,-1) {$t$};

	\draw (1) to (2);
	\draw (1) to (3);
	\draw (1) to (4);
	\draw (3) to (4);
	
	\draw (2) to (4);
	\draw (2) to (3);
	
	\draw (t) to (1);
	\draw (t) to (2);

\end{tikzpicture}
}
  \end{center}
%\vspace{-5mm}
\caption{$K_5$ with 2 edges incident to $t$ removed.}
	\label{fig:k5-2-poss}
\end{minipage}
\end{figure*}
%}
%\vspace{1mm}

Given our characterization of when perfect resilience is possible in a model where both the source and the destination of a packet can be matched, we now continue charting the landscape of perfect resilience by considering a model where forwarding rules can only depend on the destination.
We are able to provide an almost perfect characterization with respect to
complete and complete bipartite graphs.

\subsection{Impossibility Results}\label{subsec:dest-negative}

Foerster et al.~\cite{apocs21resilience} showed that $K_5$ and $K_{3,3}$ do not allow for perfect resilience in the destination-based model, \emph{i.e.}, $A_p(K_5,t) = \emptyset$ and $A_p(K_{3,3},t) = \emptyset$.
Their proof construction for $K_5$ starts at some node $v \neq t$, where the link  $(v,t)$ is removed, leaving all other links incident to $v$ intact.
In their construction, all nodes, except the one node connected to $t$, must route in a cyclic permutation, a fact retained even if the link  $(v,t)$ never existed.
Hence:

\begin{theorem}
%[Follows from the proof of Lemma 4.2~\cite{apocs21resilience}]
\label{thm:k5-1-does-not-work}
A complete graph with five nodes, minus one link, $K_5^{-1}$, does not allow for a perfectly resilient forwarding pattern, \emph{i.e.}, $A_p(K_5^{-1},t) = \emptyset$.
\end{theorem}

For $K_{3,3}$, Foerster et al.~\cite{apocs21resilience} start their construction on a node $v$ which is in the same part as the destination, and hence there was no link  $(v,t)$ to begin with.
However, we can observe that in their construction, the permanent loop also traverses nodes of the other part (without $t$) and the routing behavior remains unchanged if we remove one link  incident to $t$ (cyclic permutations are enforced for all non-neighbors of the destination).

\begin{theorem}
%[Follows from the proof of Lemma 4.3~\cite{apocs21resilience}]
\label{thm:k33-1-does-not-work}
A complete bipartite graph with three nodes in each part, minus one link, $K_{3,3}^{-1}$, does not allow for a perfectly resilient forwarding pattern, \emph{i.e.}, $A_p(K_{3,3}^{-1},t) = \emptyset$.
\end{theorem}

%\klaus{for Juho: \# of link  removals: I remove only 2 link, but in order to enforce permutation, a node might need to ``think'' that its neighbors can lose all further link. Not sure what that means for simulation.}

Whereas $K_{5}$ and $K_{3,3}$ are not planar, both $K_{5}^{-1}$ and $K_{3,3}^{-1}$ are planar~\cite{wagner}.
We note that $K_5^{-1}$ is a minor of the planar 7-node construction to show impossibility in~\cite[Theorem 5.3]{apocs21resilience} and hence \Cref{thm:k5-1-does-not-work} improves their planar result with a smaller number of link and nodes.

%\begin{figure}[htbp]
	%\centering
		%\includegraphics[width=0.45\textwidth]{figs/k5-apocs.png}
	%\caption{k5-1 vs apocs 7 node}
	%\label{fig:k5-apocs}
%\end{figure}

\subsubsection{Generalization of Impossibility: Minor Relationships}
It was previously shown that if a graph $G$ allows for perfect resiliency in destination-based routing, so do all minors of $G$~\cite[\S4]{apocs21resilience}. 
Hence all graphs containing a $K_{3,3}$ or a $K_{5}$ minus one link as a minor do not allow for perfect resilience.

\subsection{Possibility Results}\label{subsec:dest-positive}

\subsubsection{One Link Less Gives Perfect Resilience}

We next show that the results from \S\ref{subsec:dest-negative} are tight in the sense that removing one additional link  from these graphs allows for perfect resilience.
We will need the following result:

\begin{corollary}[Corollary 6.2~\cite{apocs21resilience}]\label{corr:extra-1-outer}
Let $G'=(V \setminus \{t\},E)$ be outerplanar.
Then $G=(V,E)$ allows for perfectly resilient forwarding patterns $\pi^t$ without the source.
\end{corollary}

\noindent We start with $K_5^{-2}$ in \Cref{thm:k5-2-does-work} and $K_{3,3}^{-2}$ in \Cref{thm:k33-2-does-work}.

\begin{theorem}\label{thm:k5-2-does-work}
A complete graph with five nodes, minus two links, $K_5^{-2}$, allows for a perfectly resilient forwarding pattern $\pi^t$, as well as for all minors of $K_5^{-2}$.
\end{theorem}

%\klaus{todo proof missing}

\begin{IEEEproof} %[Proof sketch] %(Full proof deferred to \Cref{app:dest}.)
Let the nodes of $K_5^{-2}$ be $v_1,v_2,v_3,v_4,v_5=t$.
We proceed by case distinction.
If $t$ has one or zero links removed, then the remaining 4-node graph is a proper subgraph of $K_4$ and hence is outerplanar, \emph{i.e.}, \Cref{corr:extra-1-outer} yields perfect resilience.

If $t$ has two links removed, then let w.l.o.g.\ $v_1,v_2$ be the neighbors of $t$. %, %as shown in Fig.~\ref{fig:k5-2-poss}.
Note that the graph without $t$ is a $K_4$ and hence is not outerplanar. 
In order to obtain perfect resilience, we need to visit, from the starting node $v$, all of $v_1,v_2$ being in the same component as $v$, which we can obtain by using the following forwarding pattern, where we state in the table in Fig.~\ref{fig:tab:k52} for each inport in which order outports are considered (using the table notation introduced in the proof of Theorem~\ref{thm:k33_with_source}):

The correctness of our algorithm follows by careful case distinction, showing that $v_1$ or $v_2$ will be visited.
%proof of our algorithm is deferred to \Cref{app-sub-dest-p} due to space constraints.
Lastly, the proof extends to all minors of $K_{5}^{-2}$ due to~\cite[Thm 4.3]{apocs21resilience}.
\end{IEEEproof}

%\begin{figure}[htbp]
	%\centering
		%\includegraphics[width=0.45\textwidth]{figs/k5-2-outer.png}
	%\caption{k5-2 outer}
	%\label{fig:k5-2-outer}
%\end{figure}

\begin{theorem}\label{thm:k33-2-does-work}
A complete bipartite graph with three nodes in each part, minus two links, $K_{3,3}^{-2}$, allows for a perfectly resilient forwarding pattern $\pi^t$, as well as for all minors of $K_{3,3}^{-2}$.
\end{theorem}

\begin{IEEEproof}
Denote nodes of the first part as $v_1,v_2,v_3=t$ and second part as $v_4,v_5,v_6$.
If $v_3=t$ has zero or one link removed, then the remaining 5-node graph is a proper subgraph of $K_{2,3}$ and hence is outerplanar, \emph{i.e.}, we obtain perfect resilience with \Cref{corr:extra-1-outer}.
If $t$ has two link removed, then there is only one node connected to $t$, w.l.o.g.\ $v_6$, and the graph without $t,v_6$ is a $K_{2,2}$, which is outerplanar. 
We hence obtain perfect resilience by first routing to $v_6$ with \Cref{corr:extra-1-outer} and then to $t$.
The proof extends to all minors of $K_{3,3}^{-2}$~\cite[Theorem 4.3]{apocs21resilience}.
\end{IEEEproof}

%\clearpage %remove for final version

\section{Resilience with Few Failures}\label{sec:few-removals}

We now study failover routing given a promise that only a small fraction of link are removed. It should be noted that in general we can use any constant-sized graph that does not have a perfectly resilient forwarding pattern, and pad it with extra unhelpful link (and nodes) such that the fraction of link that fail can be made arbitrarily small. Therefore the general case is uninteresting, and we must consider specific graph classes.

We consider routing with source and destination information on complete graphs and complete bipartite graphs. In this setting, there are no perfectly resilient forwarding patterns for $K_7$ 
%(Lemma~\ref{nok7}) 
and $K_{4,4}$ 
%(Lemma~\ref{lem:nok44_with_source})
(\S\ref{subsec:source-negative}). We use a simulation argument to extend these results to graphs of any size: 
%complete graphs $K_n$ and complete bipartite graphs $K_{a,b}$ do not have forwarding patterns even if only $O(\sqrt{|E|})$ respectively $O(a+b)$ links fail. 
complete graphs and complete bipartite graphs  do not have forwarding patterns even if only $O(\sqrt{|E|})$ links fail. 
In the context of routing without source information slightly better constants can be achieved using the constructions of Foerster et al.\ \cite{apocs21resilience}.

\begin{theorem} \label{thm:complete-edge-removal}
  For every forwarding pattern on the complete graph $K_n$ on $n \geq 8$ nodes, there is a set of link failures of size at most $6n - 33$ such that the forwarding pattern fails.
\end{theorem}

\begin{IEEEproof}
  Assume that the claim does not hold for some $n$. Then there must exist a forwarding pattern $\pi^{s,t}$ that succeeds even if any $r(n)$ links fail. We simulate $\pi^{s,t}$ on the complete graph $K_7$ and reach a contradiction with the impossibility of perfectly resilient routing on~$K_7$~(\S\ref{subsec:source-negative}).
	%from Lemma~\ref{nok7}.
  %
  Given $K_7$, construct a virtual $K_n$ by adding $n-7$ virtual nodes and virtual links between all pairs of nodes. We construct a failure pattern for $K_n$ that contains the real failure pattern on $K_7$ as a subset, and simulate $\pi^{s,t}$. The failure set $F$ on $K_n$ is defined as follows.
  \begin{enumerate}
    \item Fail all links between the non-destination nodes of $K_7$ and the virtual nodes ($6(n-8)$ links in total).
    \item Fail all links that can fail in $K_7$ ($\leq 15$ links by Corollary~\ref{corr:edge-removal-k7}). 
  \end{enumerate}
  We do not need to fail any additional links incident to the destination.
  Each node $v$ in $K_7$ can use the forwarding function $\pi^{s,t}_v$ as if it were on $K_n$. Since only the destination is connected to the virtual nodes, the packet will never leave the real network $K_7$. Since we assumed $\pi^{s,t}$ forwards correctly on $K_n$, it must also forward correctly on this particular failure set. Therefore it forwards correctly on $K_7$, a contradiction.
  In total, the number of links in $F$ is $6(n-8) + 15 = 6n - 33$.
\end{IEEEproof}

The result is asymptotically the best possible. For example Foerster et al.\ showed that forwarding with source and destination is always possible if the distance between $s$ and $t$ in $G \setminus F$ is at most 2~\cite[Theorem 6.1]{apocs21resilience}. This holds on the complete graph when $|F| \leq n-2$.

We can give a similar construction for complete bipartite graphs. The proof is a simulation argument based on the impossibility of the  $K_{4,4}$ (\S\ref{subsec:source-negative}):
%Lemma~\ref{lem:nok44_with_source}

\begin{theorem} \label{thm:bip-complete-edge-removal}
  For every forwarding pattern on the complete bipartite graph $K_{a,b}$, where $a \geq b \geq 4$, there is a set of link failures of size $3a + 4b - 21$ s.t.\ the forwarding pattern fails.
\end{theorem}

\begin{IEEEproof}
  Consider an instance of $K_{4,4}$ such that the node that has the packet initially is on the same side as the destination. Create the virtual graph $K_{a,b}$ by adding $a+b-8$ nodes and the corresponding links. If $a > b$, we let the larger part be on the opposite side of the source and the destination. 
  
  The failure set $F$ is the union of the following sets.
  \begin{enumerate}%[noitemsep] 
    \item The real failure set of $K_{4,4}$ (in total 11 links by Cor.~\ref{corr:edge-removal-k44}).
    \item All links from the non-destination nodes of $K_{4,4}$ to the virtual nodes (in total $3(a-4) + 4(b-4)$ links). 
  \end{enumerate}
  
  Again we can simulate the forwarding pattern $\pi^{s,t}$ for $K_{a,b}$ in $K_{4,4}$ since the packet will never enter the virtual part of the graph. Assuming the packet is forwarded correctly on $K_{a,b}$, it is forwarded to the destination on the subgraph that corresponds to $K_{4,4}$, a contradiction (\S\ref{subsec:source-negative}).
	%with Lemma~\ref{lem:nok44_with_source}.
  The total size of the failure set $F$ is at most $3a + 4b - 21$. 
\end{IEEEproof}

Chiesa et al.~\cite[\S B.2, B.3]{Chiesa2014} showed how to survive $k-1$ link failures in $k$-connected complete and complete bipartite graphs. This implies that our result is asymptotically best possible on balanced complete bipartite graphs.

We further briefly investigate the transfer of resilience under few failures to subgraphs and minors.
More precisely, if a graph $G$ allows for a $k$-resilient forwarding pattern, do the subgraphs (respectively, minors) of $G$ then also allow for a $k$-resilient forwarding pattern?
This property does not hold in general.
We know that, \emph{e.g.}, the $K_{100}$ is 99-connected and thus $98$-resilient~\cite[\S B.2]{Chiesa2014}.
On the other hand, we know that $K_7$ is not perfectly resilient (\S\ref{subsec:source-negative}),
%, by \Cref{nok7}, 
and $K_7$ is a subgraph of the $K_{100}$. 
However, $|E(K_7)|<98$, and hence 98-resilience is equivalent to perfect resilience on $K_7$, \emph{i.e.}, the $98$-resilience of $K_{100}$ does not carry over to its subgraphs.
An analogous statement can be made, \emph{e.g.}, with $K_{100,100}$, applying~\cite[\S B.3]{Chiesa2014} and impossibility of perfect resilience on $K_{4,4}$ (\S\ref{subsec:source-negative}), %~\Cref{lem:nok44_with_source}.
As subgraphs are also minors, we hence answer the question in the negative for both complete and complete bipartite~graphs. 

\section{From Routing to Touring:
Perfect Resilience Without Source and Destination}\label{sec:touring}

While we have so far focused on the standard routing problem, namely
delivering a packet from the source to the destination, in this section, we extend our investigations to a fundamental touring problem: Can local rerouting rules be defined which ensure that a packet will visit \emph{all} nodes in a graph, even under failures? 
At first this problem seems quite different to normal routing, but touring and routing are deeply connected on complete graphs: in order to reach the destination $t$, we need to tour all of its neighbors, as an adversary could disconnect $t$ from all neighbors except one.

Our results match the above intuition: as we will show in this section, the borders of (im)possibility move by exactly one node between touring and destination-based routing on complete graphs.
What's more, we will provide a complete classification of touring under perfect resilience in \Cref{corr:perfect-touring}.

Beyond the above theoretical motivation, touring can also help in a practical context, by saving expensive routing table space: we deploy the same routing rules, no matter which source or destination a packet has.
First, for destination-based routing, the packet will eventually reach the destination, and can then be removed from the network.
Second, if we also have the source, we can use touring to implement a broadcast or flooding protocol.
Once the source gets the packet again, it checks if the next outport is the same outport as for $\perp$: if yes, the packet has toured the whole network (assuming resilience), and if not, it is still underway in its tour.

%For the setting where we need to \emph{tour} all nodes in the current component, without a source or destination, we will use the analogous notation of $\pi_v^{\forall}(e,F)$ resp.~$\pi_v^{\forall}(u,F)$.

\subsection{Complete Touring Characterization in Perfect Resilience}\label{subsec:class-tour}

We will now present a complete characterization of touring.
Let us first introduce some terminology. 
We will denote a forwarding pattern $\pi^{\forall}$ as \emph{$k$-resilient} if for all $G$ and all $F$ where $|F| \leq k$ the forwarding pattern $\pi^{\forall}$ routes the packet from all $v \in V$ to all nodes $v'$ in the connected component of $v$ in $G\setminus F$ and then back to $v$.
We call a forwarding pattern $\pi^{\forall}$ \emph{perfectly resilient} if it is $\infty$-resilient: all nodes are visited in the tour through the connected component.
Let $A_p(G,\forall)$ be the set of such perfectly resilient touring patterns (algorithms).%, respectively $A_p(G)$ or $A_p$ when the context is~clear.

%
%We next provide a complete classification of touring under perfect resilience.
We can now state our main technical result of this section, yielding a complete classification in \Cref{corr:perfect-touring}. %Some proofs in this part are deferred to \Cref{app:tour} due to space constraints.

\begin{theorem}\label{thm:all-non-outerplanar-no}
If $G$ is not outerplanar, then it does not support a perfectly resilient touring pattern $\pi^\forall$. %, i.e., $A_p(G) = \emptyset$.
\end{theorem}

It follows from the arguments of Foerster et al.~\cite[\S 6.2]{apocs21resilience} that every outerplanar graph can be toured, by providing a planar embedding and routing according to the right-hand rule, starting on the outer face. 
In combination with Theorem~\ref{thm:all-non-outerplanar-no} we hence obtain a complete classification of the possibility of touring all nodes:
%In other words:

%\begin{theorem}[\S 6.2~\cite{apocs21resilience}]\label{thm:apocs-outer}
%Every outerplanar graph allows for a perfectly resilient touring pattern $\pi^{\forall}$.
%\end{theorem}

%Combining Theorems~\ref{thm:all-non-outerplanar-no} and~\ref{thm:apocs-outer} hence provides a complete classification of the possibility of touring all nodes:

\begin{corollary}\label{corr:perfect-touring}
A graph $G$ allows for a perfectly resilient touring pattern if and only if $G$ is outerplanar.
\end{corollary}

It remains to prove \Cref{thm:all-non-outerplanar-no}.
To this end, we first state the auxiliary \Cref{thm:all-permutation} which we use to show that $K_4$ and $K_{2,3}$ do not allow for a perfectly resilient forwarding pattern, its correctness follows analogously as for~\cite[Lemma 3.1]{apocs21resilience}:
%

%\review{[Section 6] Please improve the writing of Lemma 1, very confusing.}

\begin{lemma}\label{thm:all-permutation}
Let $G=(V,E)$ with $|E|>0$ and let $A \in A_p(G,\forall)$, i.e., $A$ is a perfectly resilient touring pattern.
For all $F$ holds:
%and all $i \in V$ 
%it holds: all neighbors of $i$ must be part of the same orbit in $A$'s forwarding function $\pi_i^\forall(\cdot,F)$ of $i$, \emph{i.e.}, 
every node routes under $A$ according to a cyclic permutation of all its neighbors, no matter the failure set $F$.
\end{lemma}

As $K_4$ and $K_{2,3}$ are the forbidden minors of outerplanar graphs, we can then show that no non-outerplanar graph can be toured under perfect resilience.
%
%Combined with the fact that the right-hand rule in outerplanar graphs can be used for perfectly resilient touring, we thus obtain that a graph can be toured perfectly if and only if it is outerplanar.
%\juho{Is the $A_{\forall}$-notation defined?} \klaus{now defined}
%\juho{Should the statement clarify that this is cyclic after failures?}\klaus{clarified it}
%
%\begin{IEEEproof}
%The statement holds immediately for degree 1 nodes, as the packet must bounce back. 
%%
%Hence, we consider graphs with at least 2 links and 3 nodes, and only investigate nodes with at least two neighbors after failures (where the failure set can also be empty).
%Let $v$ be such a node with neighbors $v_1,\ldots,v_k$, $k \geq 2$ after failures.
%%
%Fail all surviving links that are not incident to $v$, meaning that the local view of $v$ stays unchanged. 
%%
%Consider a packet that starts its tour at $v_1$, then it must visit all neighbors of $v$ in some order and then return to $v_1$, \emph{e.g.}, $v_1-v-v_2-v-v_3-\ldots-v_k-v-v_1$. This is impossible without $v$ routing according to a cyclic permutation of all its neighbors.
%\end{IEEEproof}
%\juho{Does this proof implicitly assume that the graph can become disconnected, and the packet only has to tour the connected component?}\klaus{yes and clarified in model}
%
We next study the forbidden minors of outerplanar graphs, as first described by Chartrand and Harary~\cite[Thm.~1]{AIHPB_1967__3_4_433_0}, which we restate as \Cref{thm:outer-minor}: % and then show impossibilites:

\begin{lemma}\label{thm:outer-minor}
A graph $G$ is outerplanar if and only if it contains no $K_4$ or $K_{2,3}$ as a minor.
\end{lemma}

The arguments for the next lemmas follow analogously as for \Cref{thm:k33-1-does-not-work,thm:k5-1-does-not-work}, leveraging the fact that in order to reach the destination therein, all other nodes need to be visited.

\begin{lemma}\label{thm:all-k4-no}
The complete graph $K_4$ with four nodes does not support a perfectly resilient touring pattern $\pi^\forall$.
%to be toured under perfect resilience, i.e.,~$A_p(K_4,\forall) = \emptyset$.
\end{lemma}
\begin{lemma}\label{thm:all-k23-no}
The complete bipartite graph $K_{2,3}$ with five nodes, two in one part and three in the other, 
does not support a perfectly resilient touring pattern $\pi^\forall$.
%does not allow to be toured under perfect resilience, i.e., $A_p(K_{2,3}) = \emptyset$.
\end{lemma}

%
%
%%
%\begin{IEEEproof}
%%
%Let $V(K_{2,3})=\{v_1,v_2,v_3,v_4,v_5\}$, with the first part containing $v_1$ and $v_2$. Assume that the packet starts in the first part, w.l.o.g.\ at $v_1$.
%%
%Assume due to Theorem~\ref{thm:all-permutation}, again w.l.o.g., that $v_1$'s cyclic forwarding permutation is $v_5v_4v_3$, sending to $v_3$ with inport $\perp$.
%%
%We now fail the link $(v_2,v_5)$, as shown in Fig.~\ref{fig:k23-imposs}.
%%
%As the cyclic forwarding permutation of $v_3$ is $v_1v_2$, the cyclic forwarding permutation of $v_4$ is $v_1v_2$, and the cyclic forwarding permutation of $v_2$ is $v_3v_4$ (\Cref{thm:all-permutation}), the routing gets stuck in the loop $v_1-v_3-v_2-v_4-v_1$ ($v_1$ routes packets from $v_4$ to $v_3$ by assumption). Hence $v_5$ is never visited and as thus $K_{2,3}$ cannot be toured under perfect resilience.
%%
%\end{IEEEproof}
%
%\klaus{number of link  removals: I remove only 1 link , but in order to enforce permutation, a node might need to ``think'' that its neighbors can lose all further link. Not sure what that means for simulation.}

Foerster et al.~\cite[\S4]{apocs21resilience} showed that perfect resilience for the destination-based model on a graph $G$ is also valid for minors $G'$ of $G$.
Their technique relies on taking a perfectly resilient forwarding pattern and showing that for the two fundamental operations in the minor relationship, namely $1)$ contracting two neighboring nodes and $2)$subsetting (taking a subgraph), that the pattern can be naturally adapted to stay perfectly resilient on the obtained minor. 
%\juho{What is a subgraphing? Should the reader know what it is?}\klaus{added explanation}
%
%The authors 
Note that a forwarding pattern can also be understood as a port mapping, where packets are forwarded from an inport to an outport, independent of source or destination, and hence the following holds:
%
%They note that considering their construction ``\textit{one can define the forwarding pattern also independent of the destination (and/or source) as a port mapping, where a packet arriving at an in-port gets forwarded to some out-port, and hence might talk simply about forwarding patterns $\pi$}''. 
%\juho{Do we really want to quote or should we paraphrase?}
%\klaus{ok, paraphrased}
%
%We state their implied result as follows:
%
\begin{corollary}
%[follows from \S4~\cite{apocs21resilience}]
\label{corr:minor-touring}
Given two graphs $G$ and $G'$ such that $G'$ is a minor of $G$, it holds that $A_p(G,\forall) \neq \emptyset$ implies that $A_p(G',\forall) \neq  \emptyset$: if $G$ permits a perfectly resilient touring pattern, so do its minors.
\end{corollary}
%\juho{Isn't this the notation for regular forwarding patterns? Why is it different from Theorem 7.1?}\klaus{good point and adapted}
%
Combining \Cref{thm:outer-minor}, \Cref{thm:all-k4-no}, \Cref{thm:all-k23-no}, and \Cref{corr:minor-touring}, we obtain the desired proof of \Cref{thm:all-non-outerplanar-no}:

%\begin{theorem}\label{thm:all-non-outerplanar-no}
%If $G$ is not outerplanar, then it does not support a perfectly resilient touring pattern $\pi^\forall$. %, i.e., $A_p(G) = \emptyset$.
%\end{theorem}
%\juho{Notation also here.}

%
\begin{IEEEproof}[Proof of \Cref{thm:all-non-outerplanar-no}.]
\Cref{thm:outer-minor} states that outerplanar graphs are characterized by not having a $K_4$ or $K_{2,3}$ as a minor.
Moreover, $K_4$ and $K_{2,3}$ do not allow perfectly resilient touring schemes according to \Cref{thm:all-k4-no,thm:all-k23-no}.
As perfect touring resiliency transfers to graph minors, due to \Cref{corr:minor-touring}, the theorem statement holds.
\end{IEEEproof}

%It follows from the arguments of Foerster et al.~\cite[\S 6.2]{apocs21resilience} that every outerplanar graph can be toured, by providing a planar embedding and routing according to the right-hand rule, starting on the outer face. In other words:
%
%\begin{theorem}[\S 6.2~\cite{apocs21resilience}]\label{thm:apocs-outer}
%Every outerplanar graph allows for a perfectly resilient touring pattern $\pi^{\forall}$.
%\end{theorem}
%
%
%Combining Theorems~\ref{thm:all-non-outerplanar-no} and~\ref{thm:apocs-outer} hence provides a complete classification of the possibility of touring all nodes:
%
%\begin{corollary}\label{corr:perfect-touring}
%A graph $G$ allows for a perfectly resilient touring pattern if and only if $G$ is outerplanar.
%\end{corollary}

\paragraph{$k$-Resilient Touring}\label{subsec:touring-hamil}
As touring is limited to outerplanar graphs, only very small complete ($K_{\leq 3}$) and complete bipartite graphs ($K_{2,2}$ and $K_{1,n}$) can be toured perfectly.
We thus also investigate touring under $k$-resilience:
\begin{theorem}\label{thm:touring-few-removals}
Let $k \in \mathbb{N}$ and let $G=(V,E)$ be a $2k$-connected complete or complete bipartite graph. 
There is a forwarding pattern $\pi^{\forall}$ s.t.\ $G$ can be toured under every $F$ with $|F| \leq k-1$.
\end{theorem}

\begin{IEEEproof}
A $2k$-connected complete or complete bipartite graph contains $k$ link-disjoint Hamiltonian cycles, following the results of Walecki~\cite{alspach2008wonderful} and Laskar and Auerbach~\cite{laskar1976decomposition}. %https://mathworld.wolfram.com/HamiltonDecomposition.html
We generate routing rules as follows, inspired by Chiesa et al.~\cite[\S B.6]{Chiesa2014}:
we enumerate the $k$ Hamiltonian cycles as $H_1,\ldots,H_k$. Starting with $H_1$, the forwarding pattern routes along $H_i$ until a failure is encountered in the next link of $H_i$ at some node $v$, upon which we switch to the next $H_j$, where $j>i$ is chosen to be minimum at $v$ (the current Hamiltonian cycle can be identified based in the incoming port).
Hence, after $k-1$ failures, at least one Hamiltonian cycle will be without failures (there are $k$ such cycles), and upon entering this Hamiltonian cycle in our routing, we continuously tour all nodes.
\end{IEEEproof}

\section{Topology Zoo Case Study}\label{sec:topo-case}

\begin{figure}[t]
  \begin{center}

  \begin{tikzpicture}[shorten >=1pt]
  \tikzstyle{vertex}=[circle,fill=black!25,minimum size=17pt,inner sep=0pt]
  \tikzstyle{edge}=[thick,black,--]

  \node[vertex] (1) at (0,0) {$v_1$};
  \node[vertex] (2) at (0,2) {$v_2$};
  \node[vertex] (3) at (2,2) {$v_3$};
  \node[vertex] (4) at (2,0) {$v_4$};
  \node[vertex] (5) at (1,-1) {$v_5$};
  \node[vertex] (6) at (1,1) {$v_6$};
  \node[vertex] (7) at (-1,1) {$v_7$};

  \foreach \x/\y in {2/3,3/4,4/1,1/6,2/6,1/7,2/7}
  {
    \draw[line width=1.2mm,purple,opacity=.3] (\x) to (\y);
  }

  \foreach \x/\y in {1/2,2/3,3/4,4/5,4/1,5/1,1/6,2/6,1/7,2/7}
  {
    \draw (\x) to (\y);
  }  

\end{tikzpicture}
\end{center}
\vspace{0.5mm}
\caption{Detailed view of the Netrail topology. This topology is not
  outerplanar: merging nodes $v_3$ and $v_4$ allows to realize the
  forbidden $K_{2,3}$ minor between $v_1,v_2$ and $v_6,v_7,v_{34}$ and
  the corresponding red edges. Hence, Touring is marked as impossible.
	However, for the destination- and source-destination-based settings, the topology is marked as sometimes: e.g., when considering $v_6$ as the destination, the remaining graph is outerplanar. Hence the neighbors of the destination can be toured.}
	%Hence, in a destination-based setting,  for destination $v_5$, perfect resilience is impossible, as removing  $v_5$ still leaves the graph non-outerplanar. However, destination-based perfect resilience is possible for all other nodes.}
	\label{fig:zooDetail1}
	%\vspace{1mm}
	\end{figure}
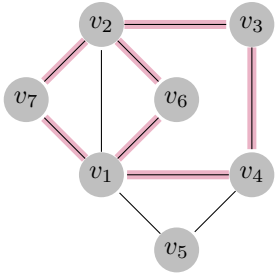
	
	\begin{table*}[]
\centering
\resizebox{\textwidth}{!}{%
\begin{scriptsize}
\begin{tabular}{>{\centering\arraybackslash}p{3.3cm}>{\centering\arraybackslash}p{1.4cm}>{\centering\arraybackslash}p{1.4cm}>{\centering\arraybackslash}p{4.1cm}>{\centering\arraybackslash}p{4.4cm}}
\textbf{Model}                    & \textbf{Subgraph}              & \textbf{Minor}
  & \textbf{Possible}                        & \textbf{Impossible}
  \\ \hline\hline 
\rowcolor{lgray}r-Tolerance: $r>1$ [\S\ref{sec:model}]        & \cellcolor{green!15}Yes [\S\ref{subsec:r-tol1}] & \cellcolor{red!15} No [\S\ref{subsec:r-tol2}]& $K_{2r+1}$ / $K_{2r-1,2r-1}$ [\S\ref{subsec:r-tol2}]                  & $K_{5r+3}$ [\S\ref{subsec:r-tol1}]                                         \\
%Source and Dest.\ [\S\ref{sec:model}]            & \cellcolor{green!15}Yes \cite[\S 4]{apocs21resilience} & \cellcolor{green!15}Yes \cite[\S 4]{apocs21resilience}     & $K_5$ / $K_{3,3}$ [\S\ref{subsec:source-positive}]                     & $K_7$ / $K_{4,4}$ minus 1 link  [\S\ref{subsec:source-negative}]                        \\
%rowcolor{lgray}Destination only [\S\ref{sec:model}]            & \cellcolor{green!15}Yes \cite[\S 4]{apocs21resilience} & \cellcolor{green!15}Yes  \cite[\S 4]{apocs21resilience}   & $K_5$ / $K_{3,3}$ minus 2 links [\S\ref{subsec:dest-positive}]             & $K_5$ / $K_{3,3}$ minus 1 link [\S\ref{subsec:dest-negative}]                     \\
Bounded \# link failures $f$ [\S\ref{sec:few-removals}] & \cellcolor{red!15}  No [\S\ref{sec:few-removals}] &
                                                             \cellcolor{red!15}  No [\S\ref{sec:few-removals}]
  & $K_n: f<n-1$ \cite[\S B.2]{Chiesa2014} \hspace{2cm} $K_{a,b}: f< \min\{a,b\}-1$ \cite[\S B.3]{Chiesa2014}                & $K_n,n>8 : f \geq 6n-33$ [\S\ref{sec:few-removals}] $K_{a,b}, a,b\geq 4: f \geq a+4b-21$ [\S\ref{sec:few-removals}]\\

%\rowcolor{lgray}No destination: Touring [\S\ref{sec:touring}]  & \cellcolor{green!15}Yes [\S\ref{subsec:class-tour}] & \cellcolor{green!15}Yes [\S\ref{subsec:class-tour}]    & $K_4$/ $K_{2,3}\not \in \textnormal{minor}(G)$:\hspace{1cm} G~outerplanar \cite[\S 6.2]{apocs21resilience} & $K_4$/$K_{2,3}\in \textnormal{minor}(G)$:\hspace{2.3cm} G not outerplanar [\S\ref{subsec:class-tour}] \\
%
\hline
\end{tabular}
\end{scriptsize}
}
\vspace{2mm}
\caption{Landscape of feasibility of local fast rerouting in different
  failure models.}
\vspace{-6mm}
\label{tbl:big1}
\end{table*}

%\todo{todo..preliminary text ahead}
%
In order to better understand the possibility of perfect resilience on real-world networks, we performed a case study on 260 networks from the Internet Topology Zoo~\cite{knight2011internet}. This data set collects information provided by network operators. The networks in this data set have between 3 and 754 nodes and between 4 and 895 links. 
We used \texttt{SageMath} 9.3\footnote{https://www.sagemath.org/} to compute if a network is outerplanar or (non-)-planar, respectively \texttt{minorminer} 0.2.6\footnote{https://github.com/dwavesystems/minorminer} to compute if it contains a forbidden minor for the respective routing model: $K_5^{-1}$ or $K_{3,3}^{-1}$ for destination-based routing and $K_7^{-1}$ or $K_{4,4}^{-1}$ for source-destination based routing. The code for this analysis has been opensourced.\footnote{https://github.com/yvonneanne/dsn22}
In case a forbidden minor was found, or if the graph was not outerplanar for touring, we marked the graph as \emph{\textcolor[rgb]{0.69,0.13,0.13}{impossible}} w.r.t. perfect resilience---on the other hand, if the network was outerplanar, we marked it as \emph{\textcolor[rgb]{0.23,0.7,0.44}{possible}}.
From the remaining graphs, if there exists a forwarding pattern for a subset of destinations\footnote{I.e, if the graph is outerplanar after removing the destination, because then we can tour all neighbors of the destination.}, we marked them as \emph{\textcolor[rgb]{0.53,0.81,0.98}{sometimes}}\footnote{We give an example in Fig.~\ref{fig:zooDetail1}.}, with the remaining networks marked as \emph{{\textcolor[rgb]{0.8,0.67,0}{unknown}}}.

\begin{figure}[t]
	\centering
		\includegraphics[width=1.00\columnwidth]{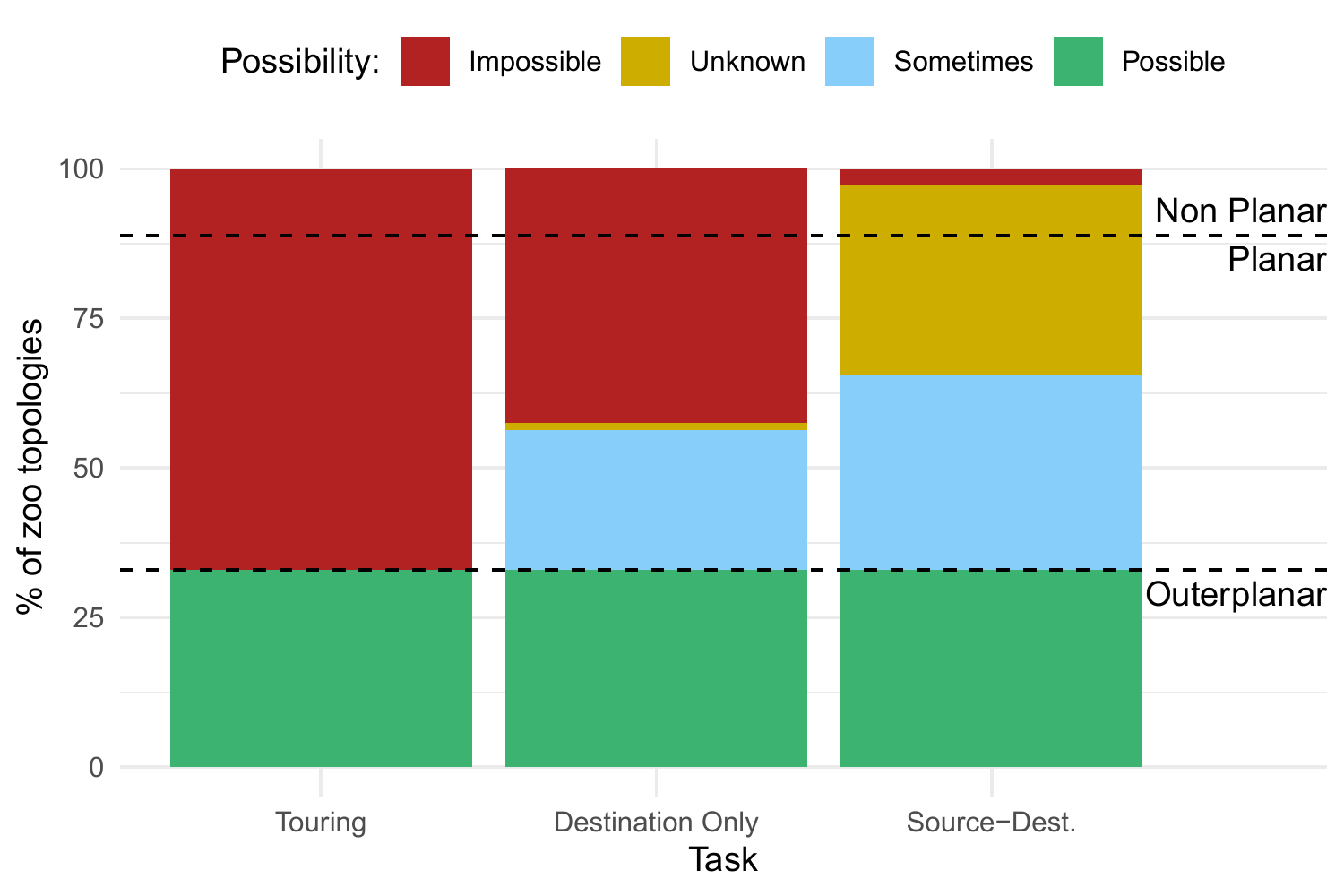}
		\vspace{-7mm}
	\caption{Perfect resilience classification of Internet Topology Zoo~\cite{knight2011internet} instances.}
	\label{fig:zooMinors}
	%\vspace{-2mm}
\end{figure}

% respectively as  if it is possible at least for a subset of destinations.
%
%The remaining networks were marked as 

The results are shown in Fig.~\ref{fig:zooMinors}.
Even though \texttt{minorminer} relies on a heuristic to solve the computationally hard minor search problem,
most instances can be classified~quickly. 

%Also note that the sparse Topology Zoo instances are particularly difficult to classify, as they are at the border between feasible and infeasible graphs. 
In general, we see that roughly one third of all topologies allow for perfect resilience.
Regarding impossibility, the remaining networks cannot be toured under perfect resilience, whereas for the other two routing models 42.5\% and 2.7\% are impossible, with 1.1\% and 31.8\% being unknown, and 23.4\% and 32.6\% allow forwarding patterns for some destinations, for routing algorithms matching on destination and source-destination, respectively.
For the topologies marked as sometimes, on average 21.3\% of the destinations are reaching perfect resilience.

%However, from the unknown topologies we can deduce that a large number allow for perfect resilience at least for some destinations, namely \todo{\%} from the 261 topologies, where we can show that we can reach on average 18.6\% of the destinations, as a lower bound.
%
%That said, as the number of unclassified topologies for source-destination-based routing is relatively large, it would be interesting to devise more rigorous (but likely slower) algorithms to provide further insights in this setting.

\begin{figure}[t]
	\centering
		\includegraphics[width=1.00\columnwidth]{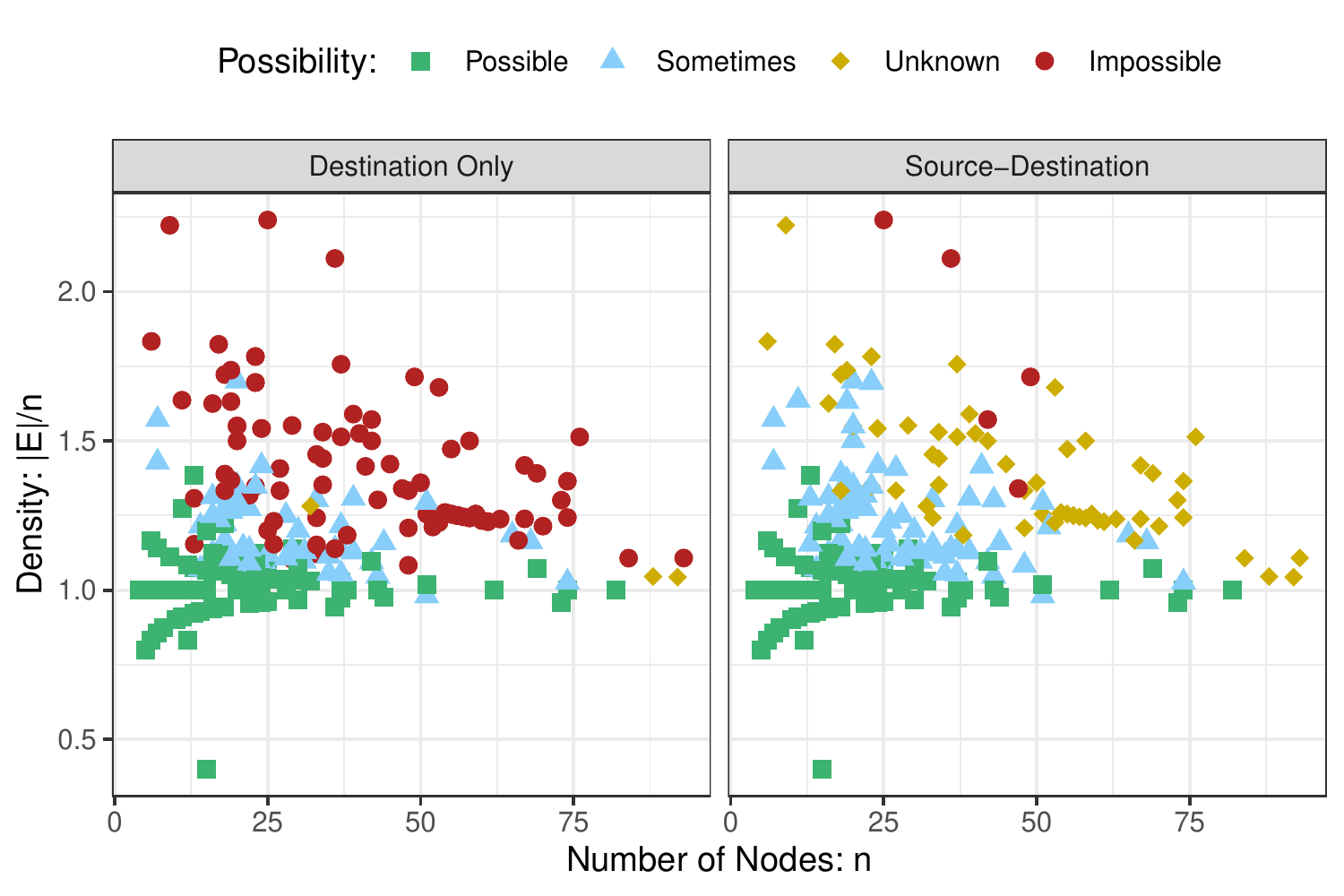}
		%\vspace{-7mm}
                \caption{Detailed view of Internet Topology
                  Zoo~\cite{knight2011internet} instances. Each topology is located by its size ($n$, x-axis) and
                  its density ($|E|/n$, y-axis). For readability,
                  large $n>100$ and dense $|E|/n>3$ outlier topologies
                  are omitted (in total 12 out of 260 topologies are omitted).}
	\label{fig:zooDetail}
	%\vspace{-2mm}
\end{figure}

%
Fig.~\ref{fig:zooDetail} presents a detailed perspective on those
results: sparse, tree-like topologies all support perfect
resilience. As density $|E|/|V|$ increases, perfect resilience becomes only
possible for some nodes (\emph{sometimes}). The densest topologies
generally do not support perfect resilience. 
For routing with source and destination the lowest density with guaranteed 
impossibility is considerably higher than for destination only routing. 
Interestingly, the impact
of density has exceptions, for instance with sparse topologies
classified as \emph{impossible} and dense topologies classified as \emph{sometimes},
confirming the importance of the local structure within each topology on
enabling perfect resilience.

Moreover, 55.8\% of all topologies are planar but not outerplanar. 
In this context the seemingly small jump in impossibility classification for destination-based routing, from $K_{5}$ or $K_{3,3}$~\cite{apocs21resilience} to $K_{5}^{-1}$ or $K_{3,3}^{-1}$ in this work, hence allows us to classify 31.3\% of the Topology Zoo instances as planar and impossible---previous work cannot show the impossibility of perfect resilience for them.
This implies that our classification measures really lie at the frontiers of (im-)possibility for the Topology Zoo data set and our new results let us classify a lot more real-world topologies.

%\review{It also analyses real-world network topologies using the developed theoretical results to classify them as based on if perfect resiliency is possible or not. Here one point could to be also comment a bit on the more practical implications of these results.}

%\rebuttal{Our analysis and code (which we will make publicly available together with the final paper) can also be used by future research to check for what further networks it would be fruitful to look for a perfectly resilient routing scheme, respectively if it is better to invest time in heuristics that work in many but not all cases. We will further emphasize and elaborate on this aspect.}
Our analysis and code can also be used by future research to check for what further networks it would be fruitful to look for a perfectly resilient routing scheme, respectively if it is better to invest time in heuristics that work in many but not all cases. If so far a destination-based routing algorithm has been used the analysis can reveal if a source-destination-based routing scheme can improve the resilience of the routing scheme. Furthermore, if network usage data is available, the most important source-destination pairs can be analysed in more detail efficiently even for very large topologies. 

%Plot 1: 86P 0U 175P 
%Plot 2: 86 P 64 U 111 I 
%Plot 3: 86 P 168 U 7 I  
%%%%%%%%%%%%%%%%%%%%%%%%%%%%%%
%3 83

%21.3%

% 88.8% planar
% 32% Outerplanar

%Btw, when we are in "sometimes" cases, we can reach on average 18.6% of the destinations.

\section{Conclusion}\label{sec:conclusion}

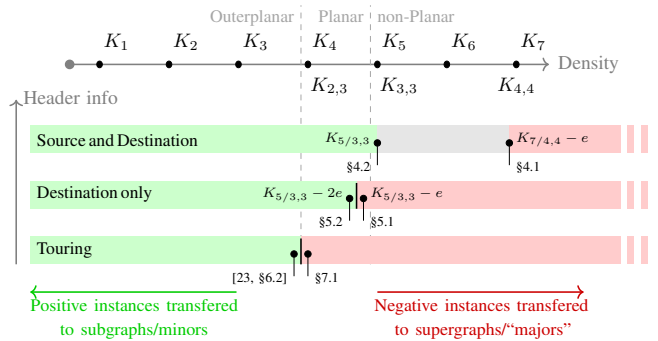
\begin{figure}[t]

  \begin{center}
\resizebox{\columnwidth}{!}{%
    \begin{tikzpicture}[shorten >=1pt,node distance=.8cm]
      \tikzstyle{vertex}=[circle,minimum size=17pt,inner sep=0pt]      
      
      \begin{scope}[xscale=1.25]
        \draw[thick,gray,Circle->] (.5,-.4) -- (7.5,-0.4)
        node[anchor=west,text width=1.5cm] {Density};

        \draw[thin,gray!70,dashed] (3.9,-4) -- (3.9,.7)
        node[anchor=north east] {\small Outerplanar};

        \draw[thin,gray!70,dashed] (4.9,-4) -- (4.9,.7)
        node[anchor=north east] {\small Planar}
        node[anchor=north west] {\small non-Planar};
        
        \foreach \i in {1,2,3,4,5,6,7} {
         \node[vertex,anchor=west] (\i) at (\i,0)  {$K_\i$};     
         \draw[fill=black] (\i,-.4) ellipse (.04 and .06);
       }

       \node[vertex,anchor=west] (k23) at (4,-.8)  {$K_{2,3}$};     
       \node[vertex,anchor=west] (k33) at (5,-.8)  {$K_{3,3}$};
       \node[vertex] (k44) at (7,-.8)  {$K_{4,4}$};
       
        \draw[thick,gray,->] (-.2,-4) -- (-.2,-1)
        node[anchor=west,text width=2.5cm] {Header info};
       
       \begin{scope}[yshift=-4cm]
         \fill[green!20!white] (0,0) rectangle (3.9,.5);
         \fill[red!20!white] (3.9,0) rectangle (8.5,.5);
         \fill[red!20!white] (8.6,0) rectangle (8.7,.5);
         \fill[red!20!white] (8.8,0) rectangle (8.9,.5);
         \draw[thick] (3.9,0)--(3.9,.5);
         
         \node[anchor=south west] (tour) at (0,0) {\small Touring};

         \draw [Circle-] (3.8, 0.25)  to (3.8, -.25) node
         [below,anchor=east] {\scriptsize [23, \S 6.2]};
         \draw [Circle-] (4, 0.25) to (4, -.25) node
         [below,anchor= west] {\scriptsize \S 7.1};
         
       \end{scope}
       \begin{scope}[yshift=-3cm]
         \fill[green!20!white] (0,0) rectangle (4.7,.5);
         \fill[red!20!white] (4.7,0) rectangle (8.5,.5);
         \fill[red!20!white] (8.6,0) rectangle (8.7,.5);
         \fill[red!20!white] (8.8,0) rectangle (8.9,.5);

         \draw[thick] (4.7,0)--(4.7,.5);
         \node[anchor=south west] (dest) at (0,0) {\small Destination\,only};

         \draw [Circle-] (4.6, 0.25)  node[anchor=east] {\scriptsize $K_{5/3,3}-2e$} to (4.6, -.25) node
         [below,anchor=east] {\scriptsize \S 5.2};
         \draw [Circle-] (4.8, 0.25) node[anchor=west] {\scriptsize $K_{5/3,3}-e$} to (4.8, -.25) node
         [below,anchor= west] {\scriptsize \S 5.1};         
       \end{scope}

       \begin{scope}[yshift=-2cm]
         \fill[green!20!white] (0,0) rectangle (5,.5);
         \fill[black!10!white] (5,0) rectangle (7,.5);
         \fill[red!20!white] (6.9,0) rectangle (8.5,.5);
         \fill[red!20!white] (8.6,0) rectangle (8.7,.5);
         \fill[red!20!white] (8.8,0) rectangle (8.9,.5);

         \node[anchor=south west] (sdest) at (0,0) {\small Source\,and\,Destination};

         \draw [Circle-] (5, 0.25)  node[anchor=east] {\scriptsize $K_{5/3,3}$} to (5, -.25) node
         [below,anchor=east] {\scriptsize \S 4.2};
         \draw [Circle-] (6.9, 0.25) node[anchor=west] {\scriptsize $K_{7/4,4}-e$} to (6.9, -.25) node
         [below,anchor= west] {\scriptsize \S 4.1};
       \end{scope}

       \begin{scope}[yshift=-4.5cm]
         \draw [thick,green!80!black, text width=5cm,align=center,
         <-] (0,0) to node[below] {\small Positive
           instances \-transfered to subgraphs/minors} (3,0);
         \draw [thick,red!80!black, text width=4.5cm,align=center, ->]
         (5,0) to node[below] {\small Negative
           instances \-transfered to supergraphs/``majors''} (8,0);
       \end{scope}

     \end{scope}
    \end{tikzpicture}
		}
  \end{center}

  %\vspace{-6mm}
\caption{Feasibility landscape of local fast rerouting in different
  routing models.}
%	\vspace{-6mm}
\label{fig:scheme1}
\end{figure}

Motivated by increasingly stringent dependability requirements,
e.g., of merging 6G communication networks, we revisited 
the algorithmic problem of realizing highly resilient fast rerouting
in the data plane. On the negative side, we proved that providing
resilience locally can be impossible, even in scenarios where the
network remains highly connected after link failures.
On the positive side, we presented improved
characterizations of resilience in various different models,
and devised novel algorithms accordingly. 
We summarize our classification results in Table~\ref{tbl:big1} and Figure~\ref{fig:scheme1}.

While our work presents a fairly complete landscape
of the achievable perfect resilience, there remain
several interesting directions for future research.
In particular, it would be interesting to chart
a similar landscape for the practically relevant scenarios in which
links failures are random, or where the routing rules
themselves can be subject to randomization.
It would further be interesting to account for additional
aspects which influence dependability in practice, and e.g.,
optimize the ``hazard value''~\cite{dsn22hazard,tacas21} of the network more generally.

\smallskip
\noindent\textbf{Acknowledgments.} 
We would like to thank our shepherd Elias P. Duarte Jr.~as well as the anonymous reviewers 
for their feedback and suggestions. 
This work was (in part) supported by the Federal Ministry of Education and Research (BMBF, Germany) as part of the 6G Research and Innovation Cluster 6G-RIC under Grant 16KISK020K,
as well as by the 
Vienna Science and Technology Fund (WWTF), project ICT19-045
(WHATIF), 2020-2024.

\smallskip
\noindent\textbf{Reproducibility.} 
Our source code will be made available at \url{https://github.com/yvonneanne/dsn22}.

\smallskip
\noindent\textbf{Bibliographical Information.} 
An extended abstract of this technical report appears at DSN'22~\cite{dsn22final}. %also ~\cite{}

%%%%%%%%%%%%%%%%%%%%%%%%%%%%%%%%%%%%%%%%%
%--Bibliography-------------------------%
%%%%%%%%%%%%%%%%%%%%%%%%%%%%%%%%%%%%%%%%%

%\clearpage
\balance

\bibliographystyle{IEEEtran} 
\bibliography{literature}

%

%\end{document}
%\clearpage
 
\begin{center}

	APPENDIX
\end{center}
%\nobalance
\section{Detailed Proofs for Section~\ref{sec:src}}\label{app-source}\nobalance
For our proofs, we will apply a corollary from \cite{apocs21resilience}, which we repeat here for the reader's convenience, along with some necessary definitions and notation.

\begin{definition}[Definition 3.1 \cite{apocs21resilience}]
  For any node $i\in G$,	$i \neq t$,	and a failure set $F$, define $F_i$ as the failures in $F$ incident to $i$, i.e., $F_i$ is the only failure set the node $i$ is aware of.
	Moreover, let $G'$ be the original graph $G$ without the links in $F_i$, i.e., $G'=G\setminus F_i$.
	A neighboring node $j \in V_{G'}(i)$ is \emph{relevant} for routing from $t$ under the failure set $F$	iff there is a path from $i$ to $t$ in $G' \setminus V'$, where $V'=V_{G'}(i)\setminus \{j\}$ is the set of all other nodes still connected to $i$.
	In other words, $j$ is a potential relay to reach $t$ from $i$'s perspective, if, in addition to $F$, all links incident to other neighbors of $i$ have failed.
  \end{definition}
	
When talking about the repeated use of a forwarding function of the node $v$ under a specific failure set $F$, we will also use the notation style~$\pi_v(\cdot, F)$, where for $a \in \mathbb{N}_{\geq 1}$, $\left(\pi^t_v(\cdot, F)\right)^a(u)$ is recursively defined as 
$$\left(\pi^{s,t}_v(\cdot, F)\right)^{(a-1)}\left(\pi^{s,t}_v(u, F)\right) =$$ $$\left(\pi^{s,t}_v(\cdot, F)\right)^{(a-2)}\left(\left(\pi^{s,t}_v(\cdot, F)\right)^2(u)\right)~\textnormal{etc.}$$
	
	\begin{definition}[Adapted from Definition 3.2 in \cite{apocs21resilience}]
Let $\pi^{s,t}_v(\cdot, F)$ be the forwarding pattern of a node $v$ for some set of failed links $F$.
We say a set of neighbors $V' \subseteq V(v)$ is in the same \emph{orbit} w.r.t.\ $\pi^{s,t}_v(\cdot, F)$, if for all pairs $v_1,v_2 \in V'$ it holds: there is some $k \in \mathbb{N}$ s.t.\ $\left(\pi^{s,t}_v(\cdot, F)\right)^k(v_1) = v_2$.
\end{definition}

\begin{corollary}\label{corr:s-orbit-ext}[Lemma 3.1 in \cite{apocs21resilience}]
  Let $G=(V,E),A\in A_p(G,s,t)$, where $s \neq  i$ is connected to $k\geq 2$ relevant neighbors $v_1,\ldots,v_k\in V$ of $i \in V$. For all $F$ where $\left| F \cap \left\{(v_1,i),\ldots,(v_k,i)\right\}\right| \leq k-2$ it holds that all relevant neighbors of $i$ under $F$ must be part of the same orbit in $A$'s forwarding function $\pi^{s,t}_i(\cdot,F)$.
\end{corollary}

\subsection{Proof of \Cref{thm:nok7-1}}

\begin{lemma}\label{nok7}
The complete graph with seven nodes does not allow for perfect resilience, i.e.,  $A_p(K_7,s,t)=\emptyset$.
\end{lemma}

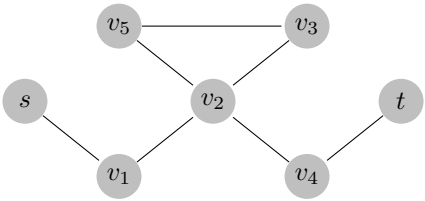
\begin{figure}[h]
  \begin{center}

  \begin{tikzpicture}[shorten >=1pt]
  \tikzstyle{vertex}=[circle,fill=black!25,minimum size=17pt,inner sep=0pt]
  \tikzstyle{edge}=[thick,black,--]

	\node[vertex] (s) at (-1.25,-1) {$s$};
	\node[vertex] (5) at (0,0) {$v_5$};
	\node[vertex] (1) at (0,-2) {$v_1$};
	\node[vertex] (2) at (1.25,-1) {$v_2$};
	\node[vertex] (3) at (2.5,0) {$v_3$};
	\node[vertex] (4) at (2.5,-2) {$v_4$};
	\node[vertex] (t) at (3.75,-1) {$t$};
	
	\draw (s) to (1);
	\draw (1) to (2);
	\draw (5) to (2);
	\draw (3) to (2);
	\draw (4) to (2);
	\draw (t) to (4);
	\draw (5) to (3);
\end{tikzpicture}
  \end{center}
\caption{$K_7$ impossibility: the nodes $v_2,v_3,v_5$ route in a cyclic permutation, due to local indistinguishability of non-local failures.}
	\label{fig:k7-imposs}
%\todo{$K_7$ has 21 link s, here only 7 remain and 14 are gone, but maybe APOCS gives better number..}
\end{figure}

The main idea of the proof is shown in Fig.~\ref{fig:k7-imposs}, where we illustrate the situation after failures.

\begin{IEEEproof}
	Let $V(K_7)= \{v_0, v_1,v_2,v_3,v_4,v_5, v_6\}$, where we assume w.l.o.g.\ $v_0$ to be the source $s$ and $v_6$ to be the destination $t$.
	To prove the lemma, we construct sets of link failures in which we leave destination $t=v_6$ connected to only one of the
  non-destination nodes, and "fine-tune" the set of link failures so that a packet emitted by $s$ will cross $v_1$ and not visit all of $v_1$'s 5 neighbors.
  By contradiction, let $A\in A_p(K_7,s,t)$. For $F_0 = \{(v_0, v_2), (v_0, v_3), (v_0, v_4), (v_0, v_6)\}$, the starting node $v_0$ is connected only to $v_1$ and $v_5$ and w.l.o.g, we assume it sends the packet to $v_1$ first. Let $\pi^{s,t}_{v_1}(\cdot,F_1)$ be the port mapping produced by $A$ at node ${v_1}$ given $F_1=\{({v_1},{v_3}),({v_1},{v_4}),({v_1},{v_5}), ({v_1},{v_6})\}$, then $\pi^{s,t}_{v_1}(v_0,F_1)=v_2$ as the only surviving path could be $s=v_0-v_1-v_2-v_6=t$.
	Since $A$ is perfectly resilient, we know by Corollary~\ref{corr:s-orbit-ext} that $\pi_{v_2}^{s,t}(\cdot, F_1)$ is a cyclic permutation over its relevant neighbors, as long as $v_2$ has at least two relevant neighbors, with $v_2$ not being connected to $s,t$.
	Hence, under $F_2 = \{(v_0,v_2),(v_6,v_2)\}$, the perfect resilient routing forms a cyclic permutation at $v_2$ and we can directly identify the predecessor of $v_1$ of $\pi^{s,t}_{v_2}(\cdot,F_2)$. It cannot be $v_1$ as then we would not have a cyclic permutation and therefore we define the predecessor to be ${v_5}$: $\pi^{s,t}_{v_2}({v_5},F_2)={v_1}$ and the successor to be ${v_5}$: $\pi^{s,t}_{v_2}({v_1},F_2)={v_3}$  w.l.o.g.
	Construct further sets of link failures as follows, $F_3=\{(v_3,v_0),(v_3,v_1),(v_3,v_4),(v_3,v_6)\}$, $F_4 = \{(v_4,v_0),(v_4,v_1),(v_4,v_3),(v_4,v_5)\}$, and lastly $F_5=\{(v_5,v_0),(v_5,v_1),(v_5,v_4),(v_5,v_6)\}$.

  Let $F= F_0 \cup F_1 \cup F_{2}\cup F_{3} \cup F_4 \cup F_5$, as shown in Fig.~\ref{fig:k7-imposs}.
	Note that from the perspective of ${v_i}$ for $i=0,1,2,3,4,5$, only the local failures are visible, and hence $F_i$ and $F$ are locally indistinguishable. Node $v_3$ has two relevant neighbors and both of them are connected to the source in $G$. The same holds for $v_5$, thus the conditions for Corollary~\ref{corr:s-orbit-ext} are satisfied and hence packets received on one port are forwarded on the other port under $F$.
	Let us follow the network traversal of a packet emitted by $v_0$. When reaching $v_1$ it is passed to $v_2$, where is must be forwarded to $v_3$ due to Corollary~\ref{corr:s-orbit-ext}.  $v_3$ sends the packet to $v_5$, then $v_5$ to $v_2$, and then $v_2$ to $v_1$ due to the same argument, where we assumed the latter due to $\pi^{s,t}_{v_2}({v_1},F_2)={v_3}$ and $\pi^{s,t}_{v_2}({v_5},F_2)={v_1}$ w.l.o.g.
	Upon arriving at $v_1$, the packet will eventually be sent back to $v_2$, either directly or via $v_1-s-v_1-v_2$, hence $A$ causes a permanent loop.
	$F$ leaves the path $s=v_0-v_1-v_2-v_4-v_6=t$ intact, and yet $A$ loops, leading to the desired contradiction.
\end{IEEEproof}

The proof arguments also hold for the graph consisting of $K_6$ and a node connected to all other nodes but one, resulting in Theorem~\ref{thm:nok7-1}.

\subsection{Proof of \Cref{corr:edge-removal-k7}}
\begin{IEEEproof}
The biggest failure set removed directly in the proof of \Cref{nok7} had 14 links, leaving only the 7 links alive shown in Fig.~\ref{fig:k7-imposs}.
However, we also called upon Corollary~\ref{corr:s-orbit-ext}, which has no restrictions on the number of links removed.
The proof of Corollary~\ref{corr:s-orbit-ext} relies on the fact that the node $i$, with $k\geq 2$ relevant neighbors, all initially connected to the source, cannot tell which of these $k$ neighbors relied the message from the source to it and which one is responsible for forwarding to the destination. 
Hence, the assumption is that the remaining graph could just consist of the $k$ links to $i$'s $k$ neighbors, and one link  from the source and one from the destination, leaving just $k+2$ links alive.
For small $k$, we can improve this bound by observing that the remaining nodes might be a disconnected component respectively where $t$ is an articulation point, but can retain the links within each other.
As thus, we require at least $k+2$ links to be alive, but the nodes not neighboring $i$ can retain all their links, except those to $s$ or neighbors of $i$.
Still, in the above proof, we applied Corollary~\ref{corr:s-orbit-ext} to node $v_2$ while node $v_2$ has $k=4$, and hereby we need to consider a failure size that leaves only 6 links alive, \emph{i.e.}, of 15.
\end{IEEEproof}

\subsection{Proof of \Cref{corr:nok44_with_source-1} and \Cref{corr:edge-removal-k44}}
%Here, we cannot directly apply Corollary~\ref{corr:s-orbit-ext}, but rather need to carefully handcraft further permutation routing arguments.

\begin{lemma}\label{lem:nok44_with_source}
The complete bipartite graph with eight nodes, four in each part, does not allow for perfect resiliency, i.e., it holds that  $A_p(K_{4,4},s,t)=\emptyset$.
\end{lemma}

\begin{IEEEproof}
  Let $V_1=\{a,b,c=t,d\},V_2=\{v_0, v_1,v_2,v_3\}$ and $E=V_1\times V_2$, where we assume w.l.o.g.\ that we start on $v_0=s$.

  By contradiction, let $A\in A_p (K_{4,4}, s, t)$.
	We will now first show, as well by contradiction, that if $(v_0,a)$ fails but the other links incident to $a$ do not fail, $a$'s forwarding pattern must be a cyclic permutation on its neighbors $v_1,v_2,v_3$.
	
	To this end, consider the failure sets 
	\begin{itemize}
		\item $F_{12} = \{(v_0,a), (v_0,c), (v_1,c), (v_2,b),(v_3,b),(v_3,c),$ $(v_0,d),(v_1,d),(v_2,d),(v_3,d)\}$,
	\end{itemize}
	 where the only $st$-path is $s=v_0-b-v_1-a-v_2-c=t$, and
	\begin{itemize}
		\item $F_{13} = \{(v_0,a), (v_0,c), (v_1,c), (v_2,b),(v_3,b),(v_2,c),$ $(v_0,d),(v_1,d),(v_2,d),(v_3,d)\}$,
	\end{itemize}
	where the only $st$-path is $s=v_0-b-v_1-a-v_3-c=t$. Hence, $\pi_{a}^{s,t}(v_1, \{(v_0,a)\})$ cannot be $v_1$ or $\perp$, but must be $v_2$ or $v_3$.	
	If $\pi_{a}^{s,t}(v_1, \{(v_0,a)\})=v_2$, then we choose the failure set $F_{13}$, which implies $\pi_{a}^{s,t}(v_2, \{(v_0,a)\})=v_3$.
	Analogously, if $\pi_{a}^{s,t}(v_1, \{(v_0,a)\})=v_3$, then we choose the failure set $F_{12}$, \emph{i.e.}, $\pi_{a}^{s,t}(v_2, \{(v_0,a)\})=v_2$.
	W.l.o.g.\ assume $\pi_{a}^{s,t}(v_1, \{(v_0,a)\})=v_2$, $\pi_{a}^{s,t}(v_2, \{(v_0,a)\})=v_3$.
	It remains to show $\pi_{a}^{s,t}(v_3, \{(v_0,a)\})=v_1$, where we immediately discard $\pi_{a}^{s,t}(v_3, \{(v_0,a)\})=\perp$.
	If $\pi_{a}^{s,t}(v_3, \{(v_0,a)\})=v_3$, we consider the failure set 
	\begin{itemize}
		\item $F_{33} = \{(v_0,a), (v_0,c), (v_1,b),(v_2,b),(v_3,c),(v_0,d),$ $(v_1,d),(v_2,d),(v_3,d)\}$,
	\end{itemize}
	 \emph{i.e.}, every $st$-path starts with $s=v_0-b-v_3-a$, and then loops due to $\pi_{a}^{s,t}(v_3, \{(v_0,a)\})=v_3$.
	If $\pi_{a}^{s,t}(v_3, \{(v_0,a)\})=v_2$, we consider the failure set 
	\begin{itemize}
		\item $F_{32}=\{(v_0,a),(v_0,c),(v_1,b),(v_2,b),(v_2,c),(v_3,c),$ $(v_0,d),(v_1,d),(v_2,d),(v_3,d) \}$,
	\end{itemize}
	 \emph{i.e.}, the only $st$-path is $s=v_0-b-v_3-a-v_1-c=t$, which must loop due to $\pi_{a}^{s,t}(v_3, \{(v_0,a)\})=v_2$ and $\pi_{a}^{s,t}(v_2, \{(v_0,a)\})=v_3$, as $v_2$ is a dead end under $F_{32}$. Hence only $\pi_{a}^{s,t}(v_3, \{(v_0,a)\})=v_1$ remains.
	Note that we can choose $F_{23}$ and $F_{22}$ analogously for $\pi_{a}^{s,t}(v_1, \{(v_0,a)\})=v_3$, $\pi_{a}^{s,t}(v_3, \{(v_0,a)\})=v_2$.
	For the remaining part of the proof, we can hence assume that when the failure set includes $(v_0,a)$ and excludes $(v_1,a),(v_2,a),(v_3,a)$, that $a$ routes according to a cyclic permutation of its neighbors, w.l.o.g.\ $(v_1,v_2,v_3)$.
	
	Note that so far, we excluded the node $d$ from our construction, but it will now play a central role.
	From $a$'s three non-source neighbors $v_1,v_2,v_3$, we will use one (w.l.o.g.\ $v_1$) to route the packet to it from $s$, one to ``hide'' the destination behind (w.l.o.g.\ $v_2$), and one to force the packet into a loop (w.l.o.g.\ $v_3$). 
	Moreover, due to the graph being bipartite, two of them require a proxy node to fulfill their goals, in the same part as $a$, for which we use the nodes $b,d$ (as $c=t$ is the destination).
	
	To this end, observe that if a node $v$ from $v_1,v_2,v_3$ has exactly two neighbors, taken from $a,b,d$, then $v$ needs to route in a cyclic permutation.
	W.l.o.g.\ let $v=v_1$ and the two surviving neighbors be $a,b$. 
	Then, we can construct failure sets s.t.\ the only surviving links are on the paths $s=v_0-a-v_1-b-v_2-c=t$ or $s=v_0-b-v_1-a-v_2-c=t$, and hence not routing in a cyclic permutation prevents perfect resilience.
	We can use similar arguments for $a,b,d$ where if they have exactly two surviving neighbors from $v_1,v_2,v_3$, then $a,b,d$ must route in a cyclic permutation: w.l.o.g.\ pick $a$ with two surviving neighbors $v_1,v_2$. 
	Here, if the only surviving path is $s=v_0-b-v_1-a-v_2-c=t$, then $a$ must forward a packet from $v_1$ to $v_2$, and if the only surviving path is $s=v_0-b-v_2-a-v_1-c=t$, then $a$ must forward a packet from $v_2$ to $v_1$, finishing this argument.
	Moreover, if a node from $a,b,d$, w.l.o.g.\ $a$, has exactly the neighbors $v_0$ and one node from $v_1,v_2,v_3$, w.l.o.g.\ $v_1$, then $a$ must forward a packet from $s=v_0$ to $v_1$, as the only surviving path could be $s=v_0-a-v_1-c=t$.
	
	Next, we consider a node $v$ from $v_1,v_2,v_3$ that has exactly the three neighbors $a,b,d$ and show that $v$, w.l.o.g.\ $v_1$, must route according to a cyclic permutation of its three neighbors, w.l.o.g.\ $(b,a,c)$. 
	To this end, observe that from the three neighbors $a,b,c$, one of them could be a dead end, one the ``relay'' from $v_0$, and one the ``relay'' to reach the destination. 
	For example,  the only surviving links could be $(v_1,b)$ and the path $s=v_0-a-v_1-d-v_3-c=t$, and we can adapt this failure pattern that each node from $a,b,c$ can play the role of dead end, ''relay'' of the source, and ``relay'' to the destination, locally indistinguishable for $v_1$, and hence $v_1$ must route in a cyclic permutation of its three neighbors.
	
	We now have all the tools to finish our proof.
	We let the surviving links be the walk $s=v_0-b-v_1-a-v_2-d-v_1-a-v_3-c=t$.
	The path is unique until hitting $v_1$, which routes in a cyclic permutation, w.l.o.g.\ $(b,a,d)$, forwarding to $a$, which routes in a cyclic permutation as well, w.l.o.g.\ $(v_1,v_2,v_3)$, forwarding to $v_2$, which forwards to $d$, which forwards to $v_1$, which now forwards to $a$ again, due to its cyclic permutation being $(b,a,d)$, now trapped in the loop $a-v_2-d-v_1-a$. 
	On the other hand, a path from $s=v_0$ to $c=t$ still exists, namely $s=v_0-b-v_1-a-v_3-c=t$, and hence the lemma statement holds by contradiction.
\end{IEEEproof}

Moreover in the proof of \Cref{lem:nok44_with_source}, the link  from the source to the destination was always considered as failed, resulting in \Cref{corr:nok44_with_source-1}.

We again briefly investigate the number of link  failures in the above proof for \Cref{corr:edge-removal-k44}.
Here we constructed the failure sets manually for each argument and did not leverage Corollary~\ref{corr:s-orbit-ext}, using at most $11$ link  failures

\subsection{Proof of \Cref{thm:k5_with_source}}\label{app:k5proof}
\begin{IEEEproof}
 We proceed by showing that packets routed with Algorithm~\ref{alg:K5_with_source} reach the destination for all possible distances between source and destination after failures.
By showing it for $K_5$, we directly show correctness for all minors of $K_5$ as well due to~\cite[Corollary 4.2]{apocs21resilience}.
 %\todo{do we need to show how to adapt the algorithm to subgraphs and minors or is there an APOCS theorem we can use?} \klaus{we can cite APOCS}
%

If the distance between source and destination is one, Line 2 of the algorithm ensures the packet arrives at its destination directly.

If the distance is two, there are four non-isomorphic candidate graphs on which  a packet could visit all other nodes before visiting $t$, $G_1, G_2, G_3, G_4$ with $V=\{s,t,x,y,z\}$ 
and link  sets $E_1=\{(x,y), (y,s), (s, z), (z,t)\}$,  $E_2=E_1 \cup \{(x,s)\}$, $E_3=\{(x,s), (s,y), (y,z), (z,t), (s,z)\}$ and $E_4=\{(s,x), (s,y), (s,z), (z,t))\}$, after removing the failed links respectively. 
Depending on how we order the IDs for $x,y,z$ for $E_1$, the algorithm may first explore $x$ before returning to $s$ but it will definitely visit $z$ via $y$ and thus find $t$.
For $E_2$, the algorithm will head straight towards $t$ if $z$ has the lowest identifier. If $y $ is the lowest identifier, the algorithm will visit the nodes in the order $s, y, x, s,z,t$ regardless of the order of the identifiers of $x,y$. 
For $E_3$, the sequence of nodes visited starts with $s, x, s$ if $x$ has the lowest identifier, followed by $y,z,t$ if $y=v$ and $z=w$ or $z,t$ otherwise. If $y$ has the lowest identifier the sequence is $s,y,z,t$, if $z=y$ it is $s,z,t$. 
For $E_4$ the algorithm guarantees that all neighbors of the source are visited if the previous ones did not connect to the destination as the nodes will send the message back if they cannot forward it to $t$.
Note that for subgraphs of $G_1, G_2, G_3, G_4$ where $(s,x)$ is missing and/or $(s,y)$ is missing from $G_4$ the destination is reached in at most the same number of steps as well by the same line of arguments, as some detours will not be taken.

If the distance is three, six non-isomorphic candidate graphs exist where a packet could visit all other nodes before visiting $t$,  $G'_1, G'_2, G'_3, G'_4, G'_5, G'_6$ with $V=\{s,t,x,y,z\}$ and link  sets $E'_1=\{(x,s), (s,y), (y,z), (z,t)\}$,  $E'_2=E_1 \cup \{(x,y)\}$, $E'_3=\{(s,x), (x,y), (y,t), (z,y)\}$, $E'_4=\{(s,x), (x,y), (y,t), (z,x))\}$, $E'_5=E_4\cup\{(z,y)\}$, and $E'_6=E_5\cup\{(z,t)\}$, after removing failed link respectively.
For $E'_1$ the algorithm will forward packets on its direct path to the destination if $y=u$. Otherwise there might be a detour to $x$ first. 
For $E'_2$,  if $x=u$ then the sequence of nodes visited is $s,x,y,z,t$, if $y=u, x=v$ then it is $s,y,x,s,x,y,z,t$, if $y=u, z=v$ or $z=u, y=v$ then no detour is taken and it the remaining case with $z=u, x=v$ the path used is $s, x, y,z,t$.
For $E'_3$, the path taken is $s,x,y,t$ and for $E_4$ a visit to $z$ might be included but no loop introduced.
For $E'_5$, $z$ is visited if $z<y$ leading to a path of $s, x, z, y,t$ and $s,x,y,t$ otherwise. 
In the last graph $E'_6$, visiting $z$ would lead to a shortcut to $t$ and in both cases $t$ is reached.
Note that for subgraphs of $G'_1$ without $(s,x) $ and $G'_3, G'_4$ without the link to $z$, the destination is reached in at most the same number of steps as well by the same line of arguments, as some detours will not be taken.

If the distance is four, the nodes form a chain and the algorithm ensures that all nodes forward the packet until it reaches it destination
\end{IEEEproof}

\subsection{Proof of \Cref{thm:k33_with_source}}\label{app:proofk33}

\begin{IEEEproof}
  We proceed similarly to the $K_5$ case. Let
  $V_1=\{a,b,c\},V_2=\{v_1,v_2,v_3\}$ and $E=V_1\times V_2$.
  
We first describe a forwarding pattern for the case where the source is not in the same part as the target and demonstrate that a packet
 forwarded accordingly reaches its destination under all failure sets if the remaining graph is connected.  We state for each node and inport combination the order in which a node tries to forward a packet to an outport if w.l.o.g.\ the source is $s=a$ and the destination is $t=v_3$: 

	%\begin{itemize}
		%\item $~s$) \quad $\perp : t, v_1,v_2$\quad\quad$v_1: v_2$ \quad\quad $v_2: v_2$ 
		%\item $~b$) \quad $v_1: t, v_2,v_1$ \quad $v_2: t, v_1,v_2$
		%\item $~c$) \quad $v_1: t, v_2,v_1$ \quad $v_2: t, v_1,v_2$
		%\item $v_1$) \quad $s: b, c, s$ \quad \quad $~~b: c, s, b$ \quad \quad $c: b, s, c$
		%\item $v_2$) \quad $s: b, c$\quad \quad\quad$~b: c, b$ \quad \quad \quad$c: b, c$
	%\end{itemize}

\vspace{1mm}
\noindent\begin{tabular}{@{}r@{\hspace{0.5cm}}r@{\hspace{0.2cm}}l@{\hspace{0.9cm}}r@{\hspace{0.2cm}}l@{\hspace{0.9cm}}r@{\hspace{0.2cm}}l}
$@ s$ & $\perp:$& $t, v_1,v_2$&$v_1:$&$ v_2$ & $v_2:$&$ v_2$ \\
$@ b$ & $v_1:$ &$t, v_2,v_1$ & $v_2:$&$ t, v_1,v_2$ & & \\
$@ c$ & $v_1:$& $t, v_2,v_1$ & $v_2:$&$ t, v_1,v_2$ & &  \\
$@ v_1$ & $s:$& $b, c, s$ &  $b:$&$ c, s, b$ &  $c:$&$ b, s, c$ \\
$@ v_2$ & $s:$& $b, c$&$b:$&$ c, b$ & $c:$&$ b, c$ \\
\end{tabular}
\vspace{1mm}

If the degree of the source after failures is three, the source has a link to the destination and the packet will be sent there directly. 
If the degree of the source after failures is two, $s$ will forward the packet to $v_1$ first. 
	\emph{Case k}: If $v_1$ is only connected to the source, the packet will be sent back to $s$ which in turn will forward it to $v_2$. 
	In this case, for source and destination to be connected  then $(v_2,b)$ and $(b,t)$
	must be up and the pattern ensures it reaches $t$. 
	\emph{Case l}: If $v_1$ has a degree of two post failures and $(v_1,x)$ with $x\in \{b,c\}$ is up, then the packet is sent to $x$. 
	If $x$ is connected to $t$ we're done, otherwise the forwarding pattern will 
	either (i) send the packet to $v_2$, if $(x,v_2)\notin F$ from where it will reach $t$ via $c$ if $x=b$ or via $b$ or (ii)
	the packet will be bouncing back to $s$ and traverse $v_2$ and the remaining node in $V_1$ connecting to $t$.
	\emph{Case m}: If all links at $v_1$ are up, the packet will first visit $b$ from where it will either (i) bounce back to $v_1$ because the degree of $b$ is one
	or (ii) reach $t$ directly or (iii) be forwarded to $v_2$ and reach $t$ via $c$. If back at $v_1$, the packet will be 
	forwarded to $c$ next and get to $t$ from there.
If the source is only connected to $v_1$ after the failures, then $v_1$ must have a remaining degree 2 or 3 and the forwarding pattern visit the nodes in $V_1$ as described in \emph{Case l .(i)} and \emph{Case m.(i)-(iii)}.
If the source is only connected to $v_2$ after the failures, then $v_2$ must have remaining degree 2 or 3, the arguments from the previous statement hold for this case too. 

Thus we have shown that for $s$ and $t$ in different parts the forwarding pattern routes a packet successfully or the failures disconnect the source from the destination.

If the source and the destination are in the same part, the following forwarding pattern allows packets to reach their destination if w.l.o.g.\ the source is $s=a$ and the destination is $t=c$ the graph remains connected under failures.

	%\begin{itemize}
		%\item $s$) \quad $\perp : v_1,v_2, v_3$ \quad $v_1: v_3, v_2$ \quad $v_2: v_3$ \quad $v_3: v_2$
		%\item $b$) \quad $v_1: v_2, v_3, v_1$ \quad $v_2: v_3, v_1, v_2$ \quad $v_3: v_1,v_2, v_3$
		%\item $v_1$) \quad $s: t, b, s$ \quad \quad $b: t, s, b$ 
		%\item $v_2$) \quad $s: t, b, s$ \quad \quad $b: t, b, s$ 
		%\item $v_3$) \quad $s: t, b, s$ \quad \quad $b: t, s, b$ 
	%\end{itemize}

\vspace{1mm}
\noindent\begin{tabular}{@{}r@{\hspace{0.2cm}}r@{\hspace{0.2cm}}l@{\hspace{0.3cm}}r@{\hspace{0.2cm}}l@{\hspace{0.3cm}}r@{\hspace{0.2cm}}l@{\hspace{0.2cm}}r@{\hspace{0.2cm}}l}
@ $s$ & $\perp :$&$ v_1,v_2, v_3$ & $v_1:$&$ v_3, v_2$ & $v_2:$&$ v_3$ & $v_3:$&$ v_2$\\ 
@ $b$ & $v_1:$&$ v_2, v_3, v_1$ & $v_2:$&$ v_3, v_1, v_2$ & $v_3:$&$ v_1,v_2, v_3$ &&\\
@ $v_1$ & $s:$&$ t, b, s$ & $b:$&$ t, s, b$ &&&&\\
@ $v_2$ & $s:$&$ t, b, s$ & $b:$&$ t, b, s$ &&&&\\
@ $v_3$ & $s:$&$ t, b, s$ & $b:$&$ t, s, b$ &&&&\\
\end{tabular}
\vspace{1mm}

Let us assume there is a failure set under which a packet will not reach the destination with this forwarding pattern.
If the degree of the source after failures is three, the packet will first go to $v_1$. 
	$(v_1,t)$  must be in the failure set, as the destination would be reached in the next hop otherwise. 
	Thus the remaining degree of $v_1$ is either one or two. In the first case the packet is sent back to $s$ from where it is sent to $v_3$.

	If $(v_3,b)$ is up, the packet is sent to $b$ and if possible forwarded to $v_2$ from where it would reach the destination. 
	Hence, $(v_2,b)$ must be down and the packet is sent back to $v_3$ from $b$ and then forwarded back to $s$ from
	where it will be sent to $v_2$ and reach the destination. 

	On the other hand, if	 $(v_1, b)$ is available, then the packet will be sent to $b$. There are four possibilities for $b$.
		If $b$ has no other neighbors the packet will go back to the source via $v_1$, then visit $v_3$ and finally reach the destination from $v_2$.
		If $b$ is connected to $v_2$ but not $v_3$ after failures, the packet will bounce back to $b$, and then reach the destination via 
		$v_1, s, v_3$.
		If $b$ is connected to $v_3$ but not $v_2$ after failures, the packet will forwarded to the destination along the sequence $b-v_3-s-v_2-t$ .
		If $(b, v_2)$ and $(b, v_3)$ are up, then the packet will reach the destination via $v_2$ if $(v_2,t)$ is 
		available or go back to $b$ and visit $v_3$ to achieve the same end result.

If the degree of the source after failures is two, $(s, x)$ and $(s, y)$ for $x, y \in V_2$ are up. 
	If the remaining degree of $x$ is one, then the packet is sent to $y$ via $s$. 
		From there it will either reach the destination directly or via $b$ and $z\in V_2\setminus\{x,y\}$ unless the graph is disconnected.
	If the remaining degree of $x$ is two, then the packet is sent to $b$ and we can distinguish between three cases for $b$. 
		If $b$ has no other neighbors the packet will go back to the source via $x$, then visit $y$.
			From there it will either reach the destination directly or via $b$ and $z\in V_2\setminus\{x,y\}$ unless the graph is disconnected.
		If $b$ is connected to $y$ but not $z$ after failures, the packet will reach the destination from $y$ unless the graph is disconnected.
		If $b$ is connected to $y$ and $z$ after failures, the packet will be forwarded to the destination as both $y$ and $z$ will visited since the forwarding pattern at $b$ is a cyclic permutation without any locally incident failures, either directly or via a detour to $s$.

	If the remaining degree of $x$ is three, then the packet is sent to $t$ directly and hence there is no failure set that causes a loop in this case. 

If the degree of the source after failures is one, the node $x\in V_2$ the packet is sent to first must have remaining degree at two, as the destination could be reached directly from there or the graph would be either disconnected otherwise. Thus the packet will be forwarded to $b$ which may be still connected to one or two nodes in $V_2$. In the former case the messages is sent to $y\in V_1, x!=y$ which must be connected to $t$ and the pacet will reach its destination. In the latter case, it will bounce back from from the next node visited and since the forwarding pattern at $b$ without any locally incident failures forms a cyclic permutation the last remaining node in $V_1$ is explored next. Thus the destination is reached in this last remaining case as well and we have demonstrated that there is no failure set that doesn't disconnect source and destination leading to a loop.

Lastly, the statement extends to all minors of $K_{3,3}$ due to~\cite[Corollary 4.2]{apocs21resilience}.
\end{IEEEproof}

\section{Detailed Proofs For Section~\ref{sec:routing-model}}\label{app:dest}

\subsection{Deferred Proof Parts for \Cref{thm:k5-2-does-work}}\label{app-sub-dest-p}
We show correctness of our algorithm, \emph{i.e.}, all of $v_1,v_2$ will be visited if possible,  by case distinction as well.

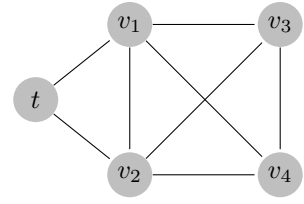
\begin{figure}[t]
  \begin{center}

  \begin{tikzpicture}[shorten >=1pt]
  \tikzstyle{vertex}=[circle,fill=black!25,minimum size=17pt,inner sep=0pt]
  \tikzstyle{edge}=[thick,black,--]

	\node[vertex] (1) at (0,0) {$v_1$};
	\node[vertex] (2) at (0,-2) {$v_2$};
	\node[vertex] (3) at (2,0) {$v_3$};
	\node[vertex] (4) at (2,-2) {$v_4$};
	\node[vertex] (t) at (-1.25,-1) {$t$};

	\draw (1) to (2);
	\draw (1) to (3);
	\draw (1) to (4);
	\draw (3) to (4);
	
	\draw (2) to (4);
	\draw (2) to (3);
	
	\draw (t) to (1);
	\draw (t) to (2);

\end{tikzpicture}
  \end{center}
\caption{Only non-outerplanar case for $K_5^{-2}$, as it is a $K_4$ when ignoring $t$. As one of the link $(v_1,t),(v_2,t)$ could fail, both $v_1,v_2$ need to be visited from any starting node in $K_4$ (if both links connected to $t$ fail, the destination is unreachable).}
	\label{fig:k5-2-poss}
\end{figure}%as shown in Fig.~\ref{fig:k5-2-poss}. 

First, assume the link  $(v_1,v_2)$ does not fail.
Then we have correctness when starting on $v_1$ or $v_2$.
When starting on $v_3$, we have correctness when $v_3$ is still neighboring $v_1$ or $v_2$, else $v_3$ is neighboring $v_4$ or $v_3$ is disconnected from all nodes.
Then, routing proceeds to $v_1$ or $v_2$ if $v_4$ is still neighboring $v_1$ or $v_2$, else $v_3,v_4$ is disconnected from $v_1,v_2$.
The argument is analogous for starting on $v_4$.

Second, assume the link  $(v_1,v_2)$ does fail.
If we start on $v_1$, we need to reach $v_2$ via $v_3,v_4$.
If $(v_1,v_3)$ is up we send to $v_3$, else to $v_4$, where we can omit the case where $v_1$ is disconnected from all neighbors.
Next, for $(v_1,v_3)$ being up, if both $(v_3,v_2)$ and $(v_4,v_1)$ are down, then the packet proceeds $v_1 - v_3-v_4-v_2$, unless $v_1,v_2$ are in separate components after failures.
Else, for $(v_1,v_3)$ being up, if $(v_3,v_2)$ is up, we reach $v_2$ from $v_1$ via $v_1-v_3-v_2$.
Lastly, for $(v_1,v_3)$ being up, if $(v_4,v_1)$ is up, but $(v_3,v_2)$ is down, then we can reach $v_2$ only via $v_4$ and if $(v_4,v_2)$ is up, else $v_1,v_2$ are in separate components after failures: if $(v_3,v_4)$ is up, via $v_1-v_3-v_4-v_2$, and if $(v_3,v_2)$ is down, via $v_1-v_3-v_1-v_4-v_2$.
For starting on $v_2$, the case is analogous and symmetrical, with $v_3,v_4$ switching places in the proof arguments.
Hence, if $v_1,v_2$ are in the same component, they reach each other.
We next cover the case of starting on $v_3,v_4$.
Again, the argument will be analogous and symmetrical for both, so hence we also only do the case distinction for $v_3$.

First, assume the link  $(v_1,v_2)$ does not fail.
If $(v_3,v_2)$ is up, then we proceed to $v_2$ and then to $v_1$ are done.
Else, if $(v_3,v_2)$ is down, we are done if $(v_3,v_1)$ is up, and else proceed to $v_4$: here, if one of $v_1,v_2$ is a neighbor of $v_4$ we are done, and else $v_1,v_2$ are not in the same component as $v_3$.

Next, assume $(v_1,v_2)$ is down.
If $(v_3,v_2)$ is up, then we proceed to $v_2$ and distinguish 2 cases.
$1)$ if $(v_2,v_4)$ is up, we go to $v_4$, where we go to $v_1$ if $(v_4,v_1)$ is up (done), else to $v_3$, if $(v_4,v_3)$ is up to $v_3$ directly and if $(v_4,v_3)$ is down, to $v_3$ via $v_4-v_2-v_3$, where we reach $v_1$, as if $(v_3,v_1)$ is not up, $v_1$ is disconnected from $v_2,v_3,v_4$.
$2)$ if $(v_2,v_4)$ is down, then we bounce back to $v_3$. If $(v_3,v_1)$ is up we are done. Else, if $(v_3,v_4)$ is up, we reach $v_4$, and as $(v_2,v_4)$ is down, we try $(v_4,v_1)$: if it is up, we are done, if it is down, then $v_1$ is disconnected from $v_2,v_3,v_4$.

Else, if $(v_3,v_2)$ is down, we consider the case whether $(v_3,v_1)$ is up.
If $(v_3,v_1)$ is up, we proceed to $v_1$, and if $v_4$ is a neighbor, we go to $v_4$: if $v_2$ is a neighbor of $v_4$ we are done, and else, $v_2$ has lost all its neighbors from $v_2,v_3,v_4$.
Should $(v_3,v_1)$ be down, then $v_3$ has a degree of 0 (done) or has $v_4$ as a neighbor, in which case we proceed to $v_4$. There, if both $v_1,v_2$ are neighbors of $v_4$, we reach them (in total) via $v_3-v_4-v_2-v_4-v_1$, and if just one of $v_1,v_2$ is a neighbor of $v_2$, then we reach that one as well (the other one is disconnected), where the case of none of $v_1,v_2$ being a neighbor of $v_4$ means that both $v_1,v_2$ are disconnected.

\section{Detailed Proofs For Section~\ref{sec:touring}}\label{app:tour}

\subsection{Detailed Proof for \Cref{thm:all-permutation}}

\begin{IEEEproof}
The statement holds immediately for degree 1 nodes, as the packet must bounce back. 
Hence, we consider graphs with at least 2 links and 3 nodes, and only investigate nodes with at least two neighbors after failures (where the failure set can also be empty).
Let $v$ be such a node with neighbors $v_1,\ldots,v_k$, $k \geq 2$ after failures.
Fail all surviving links that are not incident to $v$, meaning that the local view of $v$ stays unchanged. 
Consider a packet that starts its tour at $v_1$, then it must visit all neighbors of $v$ in some order and then return to $v_1$, \emph{e.g.}, $v_1-v-v_2-v-v_3-\ldots-v_k-v-v_1$. This is impossible without $v$ routing according to a cyclic permutation of all its neighbors.
\end{IEEEproof}

\subsection{Detailed Proofs for \Cref{thm:all-k4-no} and \Cref{thm:all-k23-no}}

\begin{IEEEproof}
Let $V(K_4)=\{v_1,v_2,v_3,v_4\}$ where we assume w.l.o.g.\ that we start on $v_1$.
Assume by \Cref{thm:all-permutation} and w.l.o.g. that $v_1$'s cyclic forwarding permutation is $v_2v_4v_3$, forwarding to $v_3$ with inport $\perp$.
Consider the failure of links $(v_2,v_3)$ and $(v_2,v_4)$, as shown in Fig.~\ref{fig:k4-imposs}.
As the cyclic forwarding permutation of $v_3$ is now $v_1v_4$, and $v_1v_3$ for $v_4$, again due to \Cref{thm:all-permutation}, the routing gets stuck in the loop $v_1-v_3-v_4-v_1$ as by assumption $v_1$ routes packets from $v_4$ to $v_3$. Node $v_2$ is never visited and as thus $K_4$ cannot be toured under perfect resilience.
\begin{figure}[t]
\begin{minipage}{0.44\textwidth}
  \begin{center}

  \begin{tikzpicture}[shorten >=1pt]
  \tikzstyle{vertex}=[circle,fill=black!25,minimum size=17pt,inner sep=0pt]
  \tikzstyle{edge}=[thick,black,--]

	\node[vertex] (1) at (0,0) {$v_1$};
	\node[vertex] (2) at (2,0) {$v_2$};
	\node[vertex] (3) at (0,-2) {$v_3$};
	\node[vertex] (4) at (2,-2) {$v_4$};

	\draw (1) to (2);
	\draw (1) to (3);
	\draw (1) to (4);
	\draw (3) to (4);
	
	\draw[dashed,red] (2) to (4);
	\draw[dashed,red] (2) to (3);
\end{tikzpicture}
  \end{center}
\caption{$K_4$ is impossible to tour.}
	\label{fig:k4-imposs}
\end{minipage}
\hfill
\vspace{10mm}
\begin{minipage}{0.44\textwidth}
\begin{center}
  \begin{tikzpicture}[shorten >=1pt]
  \tikzstyle{vertex}=[circle,fill=black!25,minimum size=17pt,inner sep=0pt]
  \tikzstyle{edge}=[thick,black,--]

	\node[vertex] (1) at (0,0) {$v_1$};
	\node[vertex] (2) at (2,0) {$v_2$};
	\node[vertex] (3) at (0,-2) {$v_3$};
	\node[vertex] (4) at (2,-2) {$v_4$};
	\node[vertex] (5) at (4,-2) {$v_5$};

	\draw (1) to (3);
	\draw (1) to (4);
	\draw (2) to (4);
	\draw (2) to (3);
	\draw (1) to (5);
	\draw[dashed,red] (2) to (5);

\end{tikzpicture}
  \end{center}
\caption{$K_{2,3}$ is impossible to tour}
	\label{fig:k23-imposs}
\end{minipage}
\end{figure}
\end{IEEEproof}
\begin{IEEEproof}
Let $V(K_{2,3})=\{v_1,v_2,v_3,v_4,v_5\}$, with the first part containing $v_1$ and $v_2$. Assume that the packet starts in the first part, w.l.o.g.\ at $v_1$.
Assume due to Theorem~\ref{thm:all-permutation}, again w.l.o.g., that $v_1$'s cyclic forwarding permutation is $v_5v_4v_3$, sending to $v_3$ with inport $\perp$.
We now fail the link $(v_2,v_5)$, as shown in Fig.~\ref{fig:k23-imposs}.
As the cyclic forwarding permutation of $v_3$ is $v_1v_2$, the cyclic forwarding permutation of $v_4$ is $v_1v_2$, and the cyclic forwarding permutation of $v_2$ is $v_3v_4$ (\Cref{thm:all-permutation}), the routing gets stuck in the loop $v_1-v_3-v_2-v_4-v_1$ ($v_1$ routes packets from $v_4$ to $v_3$ by assumption). Hence $v_5$ is never visited and as thus $K_{2,3}$ cannot be toured under perfect resilience.
\end{IEEEproof}

%---------------------------------
%% For old notes / future work / etc (to be commented out for submission)

%\include{artefacts/future-work} %future work
%\include{artefacts/old-text} %incomplete proofs

\end{document}